\documentclass[showpacs,preprintnumbers,nofootinbib]{revtex4} 

\usepackage{graphicx}
\usepackage{dcolumn}
\usepackage{bm}
\usepackage{amssymb}
\usepackage{amsmath}
\usepackage{epsfig}    
\usepackage{color}    

\def\beq{\begin{equation}}
\def\eeq{\end{equation}}
\def\bea{\begin{array}}
\def\eea{\end{array}}
\def\be{\begin{equation}}
\def\ee{\end{equation}}
\def\ba{\begin{eqnarray}}
\def\ea{\end{eqnarray}}

\def\to{\rightarrow}

\def\[{\left[}
\def\]{\right]}
\def\({\left(}
\def\){\right)}

\def\sm0{{\widetilde{m}_0}}

\def\U1em{{U(1)_{\rm em}}}
\def\to{\rightarrow}

\def\sq2{\sqrt{2}}

\def\ee{e^+e^-}

\def\End{\end{document}}

\def\fsl#1{\setbox0=\hbox{$#1$}                 
   \dimen0=\wd0                                 
   \setbox1=\hbox{/} \dimen1=\wd1               
   \ifdim\dimen0>\dimen1                        
      \rlap{\hbox to \dimen0{\hfil/\hfil}}      
      #1                                        
   \else                                        
      \rlap{\hbox to \dimen1{\hfil$#1$\hfil}}   
      /                                         
   \fi}

\begin{document} 

\title{Testing the Higgs triplet model with the mass difference at~the~LHC}%
\preprint{KANAZAWA-11-17, UT-HET 059}
\author{%
{\sc Mayumi Aoki\,$^1$, 
     Shinya Kanemura\,$^2$,   
     Kei Yagyu\,$^2$}}
\affiliation{
$^1$Institute~for~Theoretical~Physics,~Kanazawa~University, \\
  Kanazawa~920-1192, ~Japan\\
$^2$Department of Physics, University of Toyama, 3190 Gofuku, Toyama 930-8555, Japan\\}
\begin{abstract}
In general, there can be mass differences among scalar bosons of the Higgs triplet field with the hypercharge of $Y=1$. 
In the Higgs triplet model, when the vacuum expectation value $v_\Delta$ of the triplet field 
is much smaller than that $v$ ($\simeq 246$ GeV) of the Higgs doublet field as 
required by the electroweak precision data, 
a characteristic mass spectrum 
$ m_{H^{++}}^2-m_{H^+}^2\simeq m_{H^+}^2-m_{\phi^0}^2(\equiv \xi)$ appears, 
where $m_{H^{++}}$, $m_{H^+}$, $m_{\phi^0}$ are the masses of the doubly-charged ($H^{++}$), the singly-charged ($H^+$) 
and the neutral ($\phi^0=H^0$ or $A^0$) scalar bosons, respectively.  
It should be emphasized that phenomenology with $\xi\neq 0$ is drastically different from that 
in the case with $\xi=0$ 
where the doubly-charged scalar boson decays into the same sign dilepton $\ell^+\ell^+$ or the diboson $W^+W^+$ depending on 
the size of $v_\Delta$. 
We find that, 
in the case of $\xi > 0$, where $H^{++}$ is the heaviest, $H^{++}$ can be identified via the cascade decays 
such as  
$H^{++}\to H^+W^{+(*)}\to \phi^0 W^{+(*)}W^{+(*)}\to b\bar{b}\ell^+\nu\ell^+\nu$. 
We outline how the Higgs triplet model can be explored in such a case at the LHC. 
By the determination of the mass spectrum, the model can be tested 
and further can be distinguished from the other models with 
doubly-charged scalar bosons. 
\end{abstract}

\maketitle

\section{Introduction}

In spite of its crucial role to trigger the electroweak symmetry breaking, 
the Higgs sector remains unknown, and 
its essence is still mysterious. 
Exploration of the Higgs sector is 
the most important issue in current high energy physics. 
Recent results of Higgs boson searches at the Tevatron and the LHC 
have strongly constrained the mass of the Higgs boson in the Standard Model (SM) 
to be from 114 GeV to 145 GeV~\cite{LHC-Higgs}. 
The Higgs boson is expected to be discovered in near future at the LHC 
as long as the SM-like picture effectively holds.

At the same time there is no strong motivation to the 
minimal form of the Higgs sector proposed in the SM, 
where only one scalar doublet field is introduced.
In fact, extended Higgs sectors are often considered 
in the context of various scenarios for physics beyond the SM, 
in which new phenomena such as neutrino oscillation~\cite{neutrino-oscillation}, 
dark matter~\cite{DM} and/or baryon asymmetry of the Universe~\cite{BAU} 
are explained.

In order to explain tiny masses of neutrinos, 
several scenarios have been proposed in the literature, 
in which a source of 
lepton number violation is introduced with additional 
Majorana neutrinos~\cite{typeI}, a triplet scalar field~\cite{typeII} or 
triplet fermion fields~\cite{typeIII}. 
In particular, the second possibility; i.e., 
the Higgs Triplet Model (HTM), where 
a scalar triplet field with the hypercharge $Y=1$\footnote{
The convention of $Y$ is defined by the relation of $Q=I+Y$.} 
is added to the SM,   
is the simplest model which 
deduces the extended Higgs sector. 
Assuming that the triplet scalar field carries two units of lepton number, 
the lepton number conservation is violated in a trilinear interaction 
among the Higgs doublet field and the Higgs triplet field. 
Majorana masses for neutrinos are then generated 
through the Yukawa interaction of the lepton doublet and the triplet scalar field. 
When the masses of the component fields of the triplet are at the TeV scale 
or less, the model can be tested by directly detecting them, such as 
the doubly-charged ($H^{\pm\pm}$), singly-charged ($H^\pm$) and the neutral scalar bosons.

In addition to the appearance of these charged scalar bosons, 
a striking prediction of the HTM is 
the relationship among the masses of the component fields of the triplet scalar field; 
$m_{H^{++}}^2-m_{H^+}^2\simeq m_{H^+}^2-m_{\phi^0}^2(\equiv \xi)$, 
where $m_{H^{++}}$, $m_{H^+}$ and $m_{\phi^0}$ are the masses of $H^{\pm\pm}$, $H^\pm$ and $\phi^0$, respectively, 
where $\phi^0$ is $H$ or $A$ with $H$ to be the triplet-like CP-even Higgs boson and 
$A$ to be the triplet-like CP-odd Higgs boson. 
The squired mass difference $\xi$ is determined by $v$ ($\simeq 246$ GeV), 
the vacuum expectation value (VEV) of the doublet scalar field, 
and a scalar self-coupling constant. 
As such a mass difference is not forbidden by the symmetry of the model,  
we may be able to distinguish the model from the others which contain charged Higgs bosons by measuring $\xi$. 
Namely, if we discover $H^{\pm\pm}$, $H^\pm$ and $\phi^0$ and if we confirm the relationship mentioned above, 
the model could be identified.

In the previous studies, the collider phenomenology in the HTM 
has been discussed mainly by assuming $\xi=0$~\cite{gunion,delm01,dem02,4-lepton,Kadastik:2007yd,Perez:2008ha,Akeroyd-Sugiyama,Melfo:2011nx}. 
In such a case, 
$H^{++}$ decays into the same sign dilepton $\ell^+\ell^+$ 
or the diboson $W^+W^+$, depending on the size of $v_\Delta$, $m_{H^{++}}$ and also the detail of neutrino masses, 
where $v_\Delta$ is the VEV of the triplet field. 
These decay modes can be a clear signature for $H^{++}$. 
At the same time, $H^+$ decays into a lepton pair $\ell^+\nu$ 
or $W^+Z$~\footnote{Depending on $m_{H^+}$, there are the other decay modes, e.g., 
$H^+\to hW^+$, $H^+\to t\bar{b}$, etc..}. 
In the case where $H^{++}$ decays into $\ell^+\ell^+$, 
the pair production process $pp\to H^{++}H^{--}\to \ell^+\ell^+\ell^-\ell^-$ 
and the associated production process $pp\to H^{++}H^-\to \ell^+\ell^+\ell^-\nu$ 
would be useful to identify $H^{++}$ and to extract the flavor structure of the model 
at the LHC~\cite{gunion,delm01,dem02,4-lepton,Kadastik:2007yd,Perez:2008ha,Akeroyd-Sugiyama,Melfo:2011nx}. 
The upper bound of $m_{H^{++}}(\gtrsim 250-300$ GeV~\cite{LHC-H++}) has been obtained from these processes. 
In the case where $H^{++}$ decays into $W^+W^+$, on the other hand,  
the process $pp\to H^{++}H^{--}\to W^+W^+W^-W^-\to \ell^+\ell^-jjjj E_T\hspace{-4mm}/$\hspace{2.5mm}  
and $pp\to H^{++}H^-\to W^+W^+W^-Z\to \ell^+\ell^-jjjj E_T\hspace{-4mm}/$\hspace{2.5mm} 
have been studied in Ref.~\cite{Perez:2008ha}. 

However, in the case of $\xi \neq 0$, as pointed out in Refs.~\cite{4-lepton,Perez:2008ha,Akeroyd-Sugiyama,Melfo:2011nx,Chakrabarti:1998qy}, 
the situation is changed drastically as compared to that of the $\xi=0$. 
There are two cases with $\xi\neq 0$ 
depending on the sign of $\xi$. 
If $\xi$ is positive (negative), 
$H^{++}$ is the heaviest (lightest) of all the triplet-like scalar bosons. 
In the case of $\xi <0$, while $H^+$ can decay into $H^{++}W^{-(*)}$~\cite{Akeroyd-Sugiyama} 
the decay pattern of $H^{++}$ is the same as in the case of $\xi=0$.   
On the other hand, in the case of $\xi > 0$, 
the cascade decay of $H^{++}$ dominates; i.e., $H^{++}\to H^+W^{+(*)}\to\phi^0 W^{+(*)}W^{+(*)}$ 
as long as $v_\Delta$ is neither too small nor too large\footnote{Recently the importance of this cascade decay has been mentioned in Refs.~\cite{Akeroyd-Sugiyama,Melfo:2011nx}.}. 
Detailed analyses for collider signature of these processes for $\xi >0$ have not been studied so far. 

In this paper, we focus on the phenomenology of the HTM with the mass difference 
among the triplt-like scalar bosons at the LHC. 
In particular, we discuss the case with $\xi >0$. 
In this case, the limit of the mass of $H^{++}$ from the recent results at the LHC 
cannot be applied, so that the triplet-like scalar bosons with the mass of $\mathcal{O}(100)$ GeV are still allowed 
\footnote{There are parameter regions where $H^{++}$ decays into the 
same sign dilepton even in the case of $\xi \neq 0$ when $v_\Delta$ is extremely small. In such a case, 
the scenario of the triplet-like scalar boson with the mass of $\mathcal{O}(100)$ GeV is excluded by the LHC direct search. }. 
We find that $m_{H^{++}}$ may be reconstructed, for example from the process 
$pp\to H^{++}H^-\to (\phi^0W^{+(*)}W^{+(*)})(\phi^0W^{-(*)})\to (\ell^+\ell^+b\bar{b}E_T\hspace{-4mm}/\hspace{2.5mm})(jjb\bar{b})$ 
by observing the Jacobian peak in the transverse mass distribution of the $\ell^+\ell^+b\bar{b}E_T\hspace{-4mm}/\hspace{2.5mm}$ 
system. 
In addition, $m_{H^+}$ can be reconstructed, for example from the process 
$pp\to H^+\phi^0\to (\phi^0W^{+(*)})(b\bar{b})\to (\ell^+b\bar{b}E_T\hspace{-4mm}/\hspace{2.5mm})(b\bar{b})$ 
by measuring the Jacobian peak in the transverse mass distribution of the 
$\ell^+b\bar{b}E_T\hspace{-4mm}/\hspace{2.5mm}$ 
system. 
Furthermore, $m_{\phi^0}$ can be measured 
by using the invariant mass distribution of the $b\bar{b}$ system. 
From these analyses all the masses of the triplet-like scalar bosons may be able to be reconstructed. 
By measuring these mass differences,  
we may be able to distinguish models which 
contain doubly-charged scalar bosons, 
for instance, doubly-charged scalar bosons from singlet fields which are motivated by 
the Zee-Babu model~\cite{Zee-Babu} and 
that from doublet scalar bosons which are discussed in Refs.~\cite{gunion,su,AKY}.

This paper is organized as follows. 
In Sec.~II, we give a brief review of the HTM. 
In Sec.~III, the decay branching ratios of the triplet-like scalar bosons 
are evaluated in both the case of $\xi=0$ and $\xi \neq 0$. 
In Sec.~IV, we outline the mass reconstruction of 
the triplet-like scalar bosons. 
Discussions are given in Sec.~V, and 
conclusions are presented in Sec.~VI.

\section{The Higgs Triplet Model}
The Higgs sector is composed of the $Y=1$ isospin triplet scalar field $\Delta$ and 
the $Y=1/2$ doublet scalar field $\Phi$. 
The most general Higgs potential is given by 
\begin{align}
V&=m^2\Phi^\dagger\Phi+M^2\text{Tr}(\Delta^\dagger\Delta)+\left[\mu \Phi^Ti\tau_2\Delta^\dagger \Phi+\text{h.c.}\right]\notag\\
&+\lambda_1(\Phi^\dagger\Phi)^2+\lambda_2\left[\text{Tr}(\Delta^\dagger\Delta)\right]^2+\lambda_3\text{Tr}(\Delta^\dagger\Delta)^2
+\lambda_4(\Phi^\dagger\Phi)\text{Tr}(\Delta^\dagger\Delta)+\lambda_5\Phi^\dagger\Delta\Delta^\dagger\Phi. \label{pot}
\end{align}
The component fields of $\Phi$ and $\Delta$ can be parameterized as 
\begin{align}
\Phi=\left[
\begin{array}{c}
\varphi^+\\
\frac{1}{\sqrt{2}}(\varphi+v+i\chi)
\end{array}\right],\quad \Delta =
\left[
\begin{array}{cc}
\frac{\Delta^+}{\sqrt{2}} & \Delta^{++}\\
\Delta^0 & -\frac{\Delta^+}{\sqrt{2}} 
\end{array}\right],\quad \Delta^0=\frac{1}{\sqrt{2}}(\delta+v_\Delta+i\eta).
\end{align} 
Imposing the vacuum condition, we can eliminate $m^2$ and $M^2$ as
\begin{align}
m^2&=\frac{1}{2}\left[-2v^2\lambda_1-v_\Delta^2(\lambda_4+\lambda_5)+2\sqrt{2}\mu v_\Delta\right],\notag\\
M^2&=M_\Delta^2-\frac{1}{2}\left[2v_\Delta^2(\lambda_2+\lambda_3)+v^2(\lambda_4+\lambda_5)\right],\quad M_\Delta^2\equiv \frac{v^2\mu}{\sqrt{2}v_\Delta}.
\end{align}
The mass of the doubly-charged scalar bosons $H^{\pm\pm}$ $(=\Delta^{\pm\pm})$ is calculated as
\begin{align}
m_{H^{++}}^2&=M_\Delta^2-v_\Delta^2\lambda_3-\frac{v^2}{2}\lambda_5\notag\\
&\simeq M_\Delta^2-\frac{v^2}{2}\lambda_5,\quad (v^2 \gg v_\Delta^2). 
\end{align}
Mass eigenstates of the singly-charged states, CP-odd states and CP-even states are obtained by 
\begin{align}
&\left(
\begin{array}{c}
\varphi^\pm\\
\Delta^\pm
\end{array}\right)=
R(\beta_\pm)
\left(
\begin{array}{c}
w^\pm\\
H^\pm
\end{array}\right),\quad 
\left(
\begin{array}{c}
\chi\\
\eta
\end{array}\right)=
R(\beta_0)
\left(
\begin{array}{c}
z\\
A
\end{array}\right),\quad
\left(
\begin{array}{c}
\varphi\\
\delta
\end{array}\right)=
R(\alpha)
\left(
\begin{array}{c}
h\\
H
\end{array}\right),\notag\\
&R(\theta) \equiv \left(\begin{array}{cc}
\cos\theta & -\sin\theta \\
\sin\theta & \cos\theta
\end{array}\right),
\end{align}
where $w^\pm$ and $z$ are the Nambu-Goldstone bosons which are absorbed by the longitudeinal mode of $W^\pm$ and $Z$, respectively. 
The mixing angles $\beta_\pm$, $\beta_0$ and $\alpha$ are expressed as
\begin{align}
\cos\beta_\pm = \frac{v}{\sqrt{v^2+2v_\Delta^2}},\quad \cos\beta_0 = \frac{v}{\sqrt{v^2+4v_\Delta^2}},\quad
\tan2\alpha &\simeq \frac{v_\Delta}{v}\frac{4M_\Delta^2-2v^2(\lambda_4+\lambda_5)}{M_\Delta^2-2v^2\lambda_1}.
\end{align}
The mass formulae for the physical scalar bosons are 
\begin{align}
m_{H^+}^2&= M_\Delta^2\left(1+\frac{2v_\Delta^2}{v^2}\right)-\frac{1}{4}(v^2+2v_\Delta^2)\lambda_5
\simeq M_\Delta^2-\frac{v^2}{4}\lambda_5,\label{mp}
\\
m_A^2 &=M_\Delta^2\left(1+\frac{4v_\Delta^2}{v^2}\right)
\simeq M_\Delta^2,\label{ma}
\\
m_H^2 &\simeq M_\Delta^2,\notag\\
m_h^2&\simeq 2\lambda_1 v^2, 
\label{meven}
\end{align} 
for $v^2 \gg v_\Delta^2$.
Notice that Eq.~(\ref{meven}) is valid as long as $M_\Delta^2 > 2\lambda_1v^2$. 
Throughout the paper, we keep this relation. 
From above mass formulae, the mass difference $\xi$ is 
determined by $-\frac{v^2}{4}\lambda_5$. 
It is useful to define the mass difference (not squired) as $\Delta m\equiv m_{H^{++}}-m_{H^+}$.
In the limit of $v_\Delta \to 0$, Yukawa interactions and gauge interactions of $h$ become 
completely the same as those of the SM Higgs boson at the tree level.  
In this sense, we call $h$ the SM-like Higgs boson. Some theoretical bounds for the Higgs potential have been 
studied in Ref.~\cite{Arhrib:2011uy}.

The neutrino masses are generated through the Yukawa interaction;  
\begin{align}
\mathcal{L}_{\nu}&=h_{ij}\overline{L_L^{ic}}i\tau_2\Delta L_L^j+\text{h.c.},\label{yn}
\end{align}
where $h_{ij}$ is a $3\times 3$ symmetric complex matrix and 
$L_L^i$ is the $i$-th generation of the left-handed lepton doublet. 
The neutrino mass matrix is obtained as 
\begin{align}
(\mathcal{M}_\nu)_{ij}=\sqrt{2}h_{ij}v_\Delta=h_{ij}\frac{\mu v^2}{M_\Delta^2}. \label{mn}
\end{align}
By this equation, the Yukawa coupling constant $h_{ij}$ and $v_\Delta$ are related with each other. 
This characteristic 
feature is important to discuss the decay of the triplet-like scalar bosons: $H^{\pm\pm}$, $H^\pm$, $H$ and $A$. 
The decay branching fractions of these scalar bosons are discussed in the next section.  

In the HTM, the rho parameter $\rho$ is predicted at the tree level as 
\begin{align}
\rho = \frac{1+\frac{2v_\Delta^2}{v^2}}{1+\frac{4v_\Delta^2}{v^2}}\simeq 1-\frac{2v_\Delta^2}{v^2}, 
\end{align}
so that $v_\Delta$ is constrained from the current experimental data, $\rho_{\text{exp}}=1.0008^{+0.0017}_{-0.0007}$~\cite{Nakamura:2010zzi}, 
i.e., $v_\Delta\lesssim$ 8 GeV. 

We here give some comments on radiative corrections in this model. 
The relation $m_{H^{++}}^2-m_{H^+}^2 \simeq m_{H^+}^2 - m_{\phi^0}^2$ 
can be changed when radiative corrections are taken into account.
Radiative corrections in models with $\rho \neq 1$ at the tree level have
been studied in Refs.~\cite{triplet_ren1,triplet_ren2,triplet_ren3,triplet_ren4,triplet_ren5,triplet_ren6,triplet_ren7}, where the correction to
the rho parameter is mainly discussed. However, such an analysis of the radiative
correction has not been done yet for the HTM with the triplet field with $Y=1$,
neither to the rho parameter nor to the Higgs potential.  A detailed study of
radiative corrections in the HTM is an important and interesting issue which
will be performed in near future.
In this paper, we focus on the strategy of measuring the masses of the
triplet field in the case with the mass differences at the LHC, so that we
do not give a further discussion on the radiative correction. At the one-loop
level, the relation in mass differences can be rewritten as
\begin{align}
  \frac{m_{H^{++}}^2-m_{H^{+}}^2}{m_{H^+}^2-m_{\phi^0}^2} \simeq 1 + \delta_{\phi^0},\quad (\phi^0=H\text{ or }A),
\end{align}
where $\delta_{\phi^0}$ is the deviation from the tree level prediction due to radiative
corrections, which is given as a function of the masses and mixing angles. 
In principle, we may test the HTM with this kind of the corrected mass relation
instead of the tree level formula by measuring the masses of the bosons.
\section{Decay of the scalar bosons}
\begin{figure}[t]
\begin{center}
\includegraphics[width=60mm]{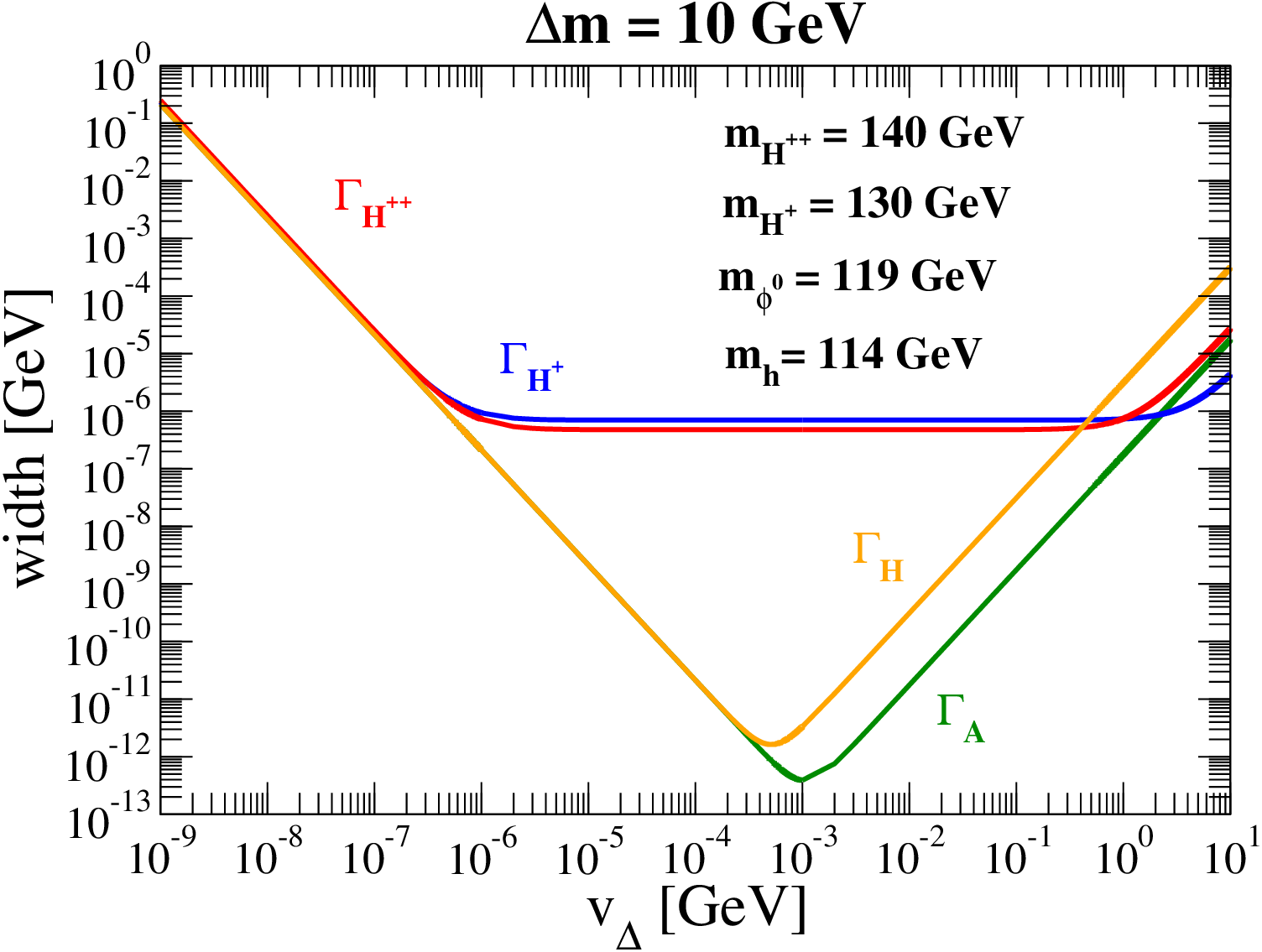}\hspace{3mm}
\includegraphics[width=60mm]{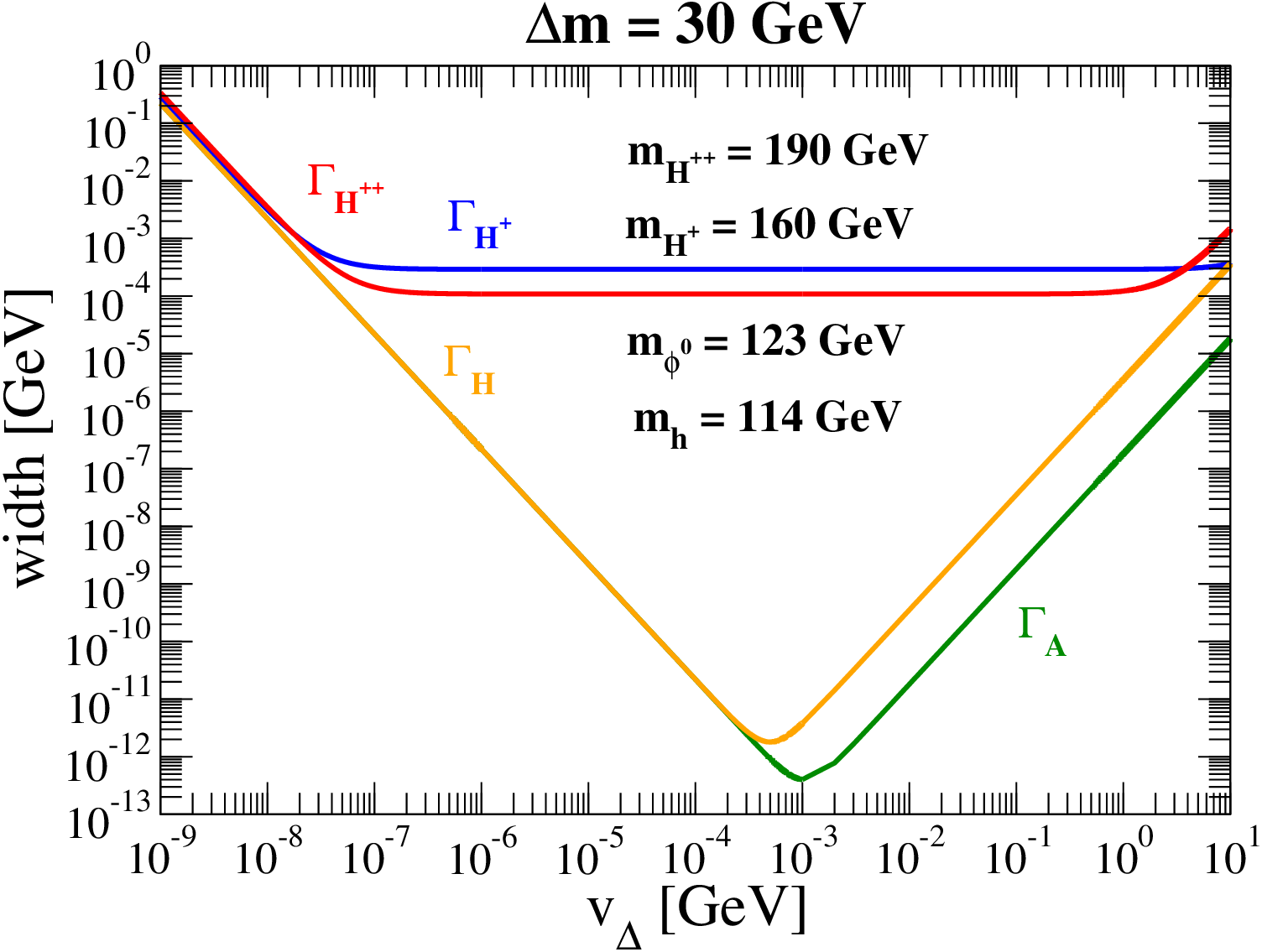}
\caption{Decay width of $H^{++}$, $H^+$, $H$ and $A$ as a function of $v_\Delta$. 
We take $m_{H^{++}}=140$ GeV (190 GeV), $\Delta m=$10 GeV (30~GeV) 
in the left (right) figure. In both the figures, $m_h$ is fixed to be 114 GeV. 
}
\label{decay-rate}
\end{center}
\end{figure}

\begin{figure}[t]
\begin{center}
\includegraphics[width=69mm]{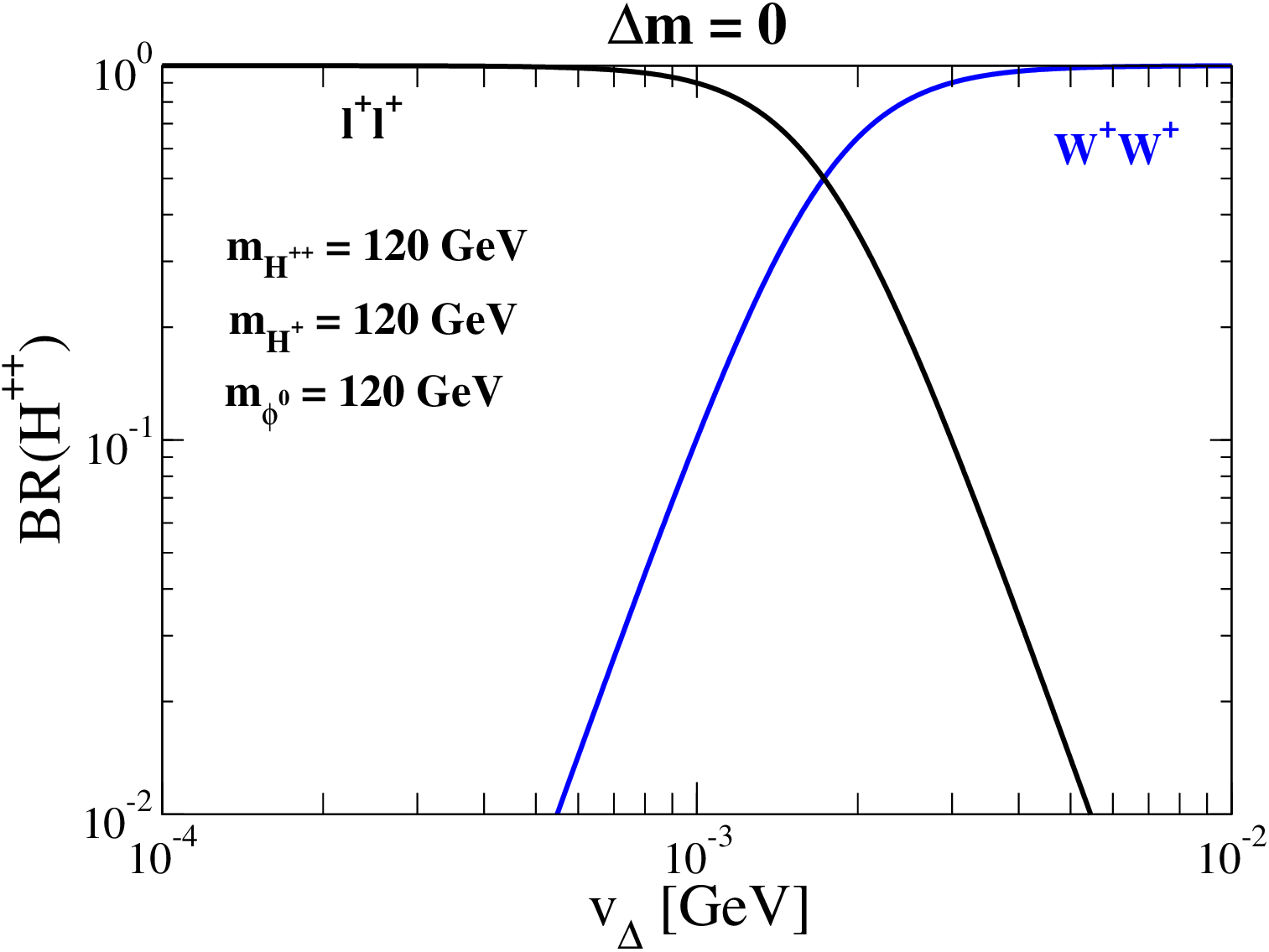}\hspace{3mm}
\includegraphics[width=65mm]{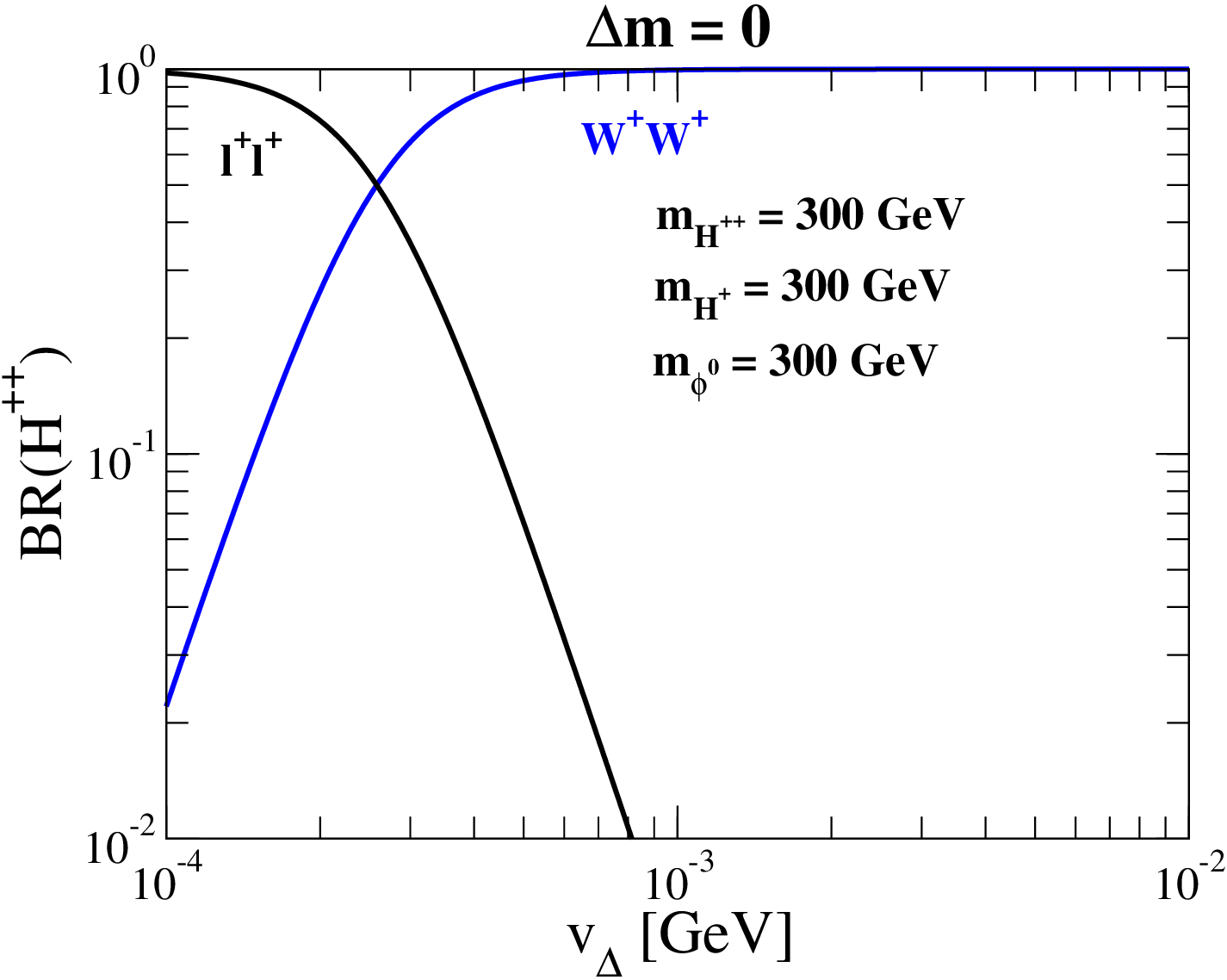}\\
\vspace{7mm}
\includegraphics[width=65mm]{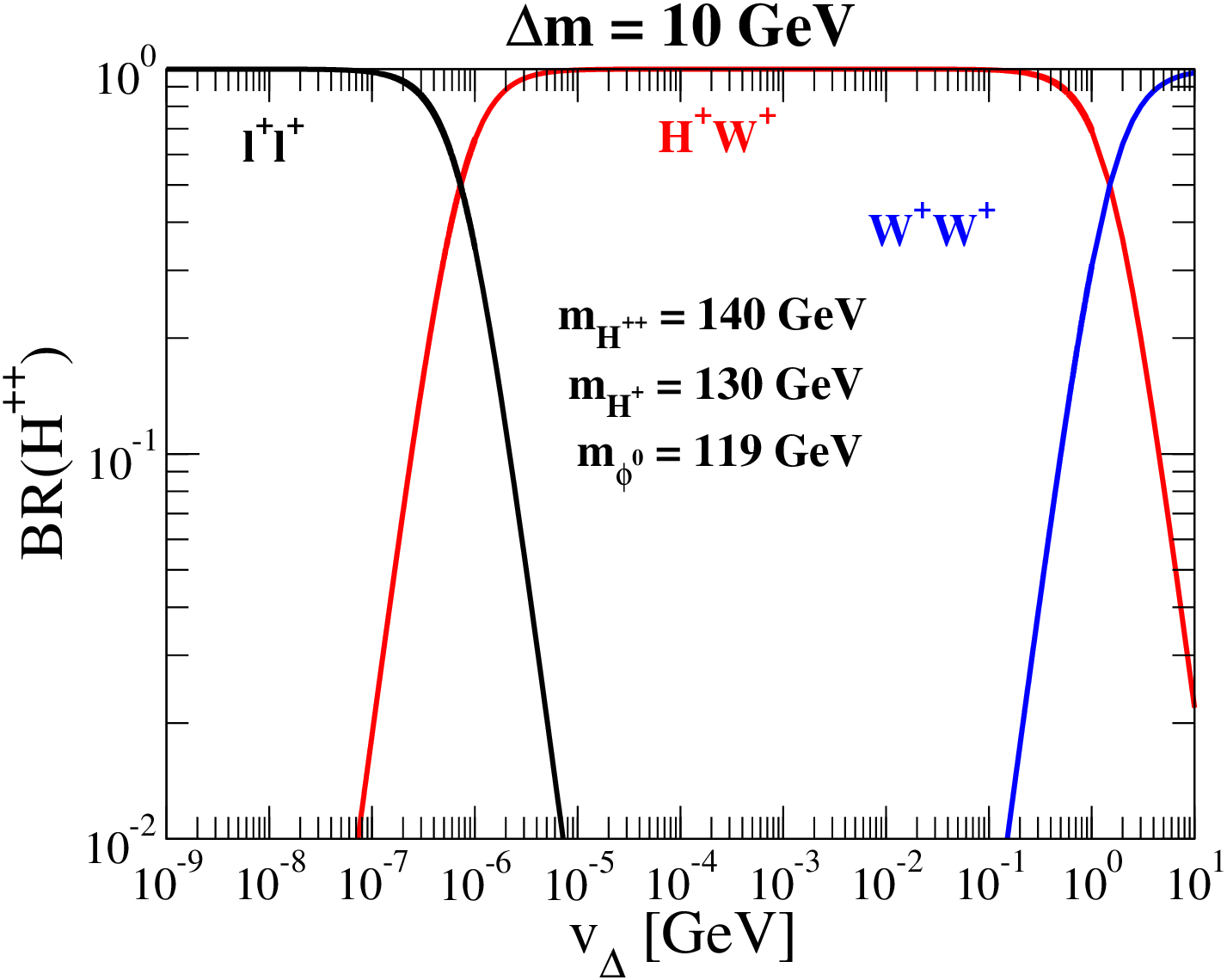}\hspace{3mm}
\includegraphics[width=65mm]{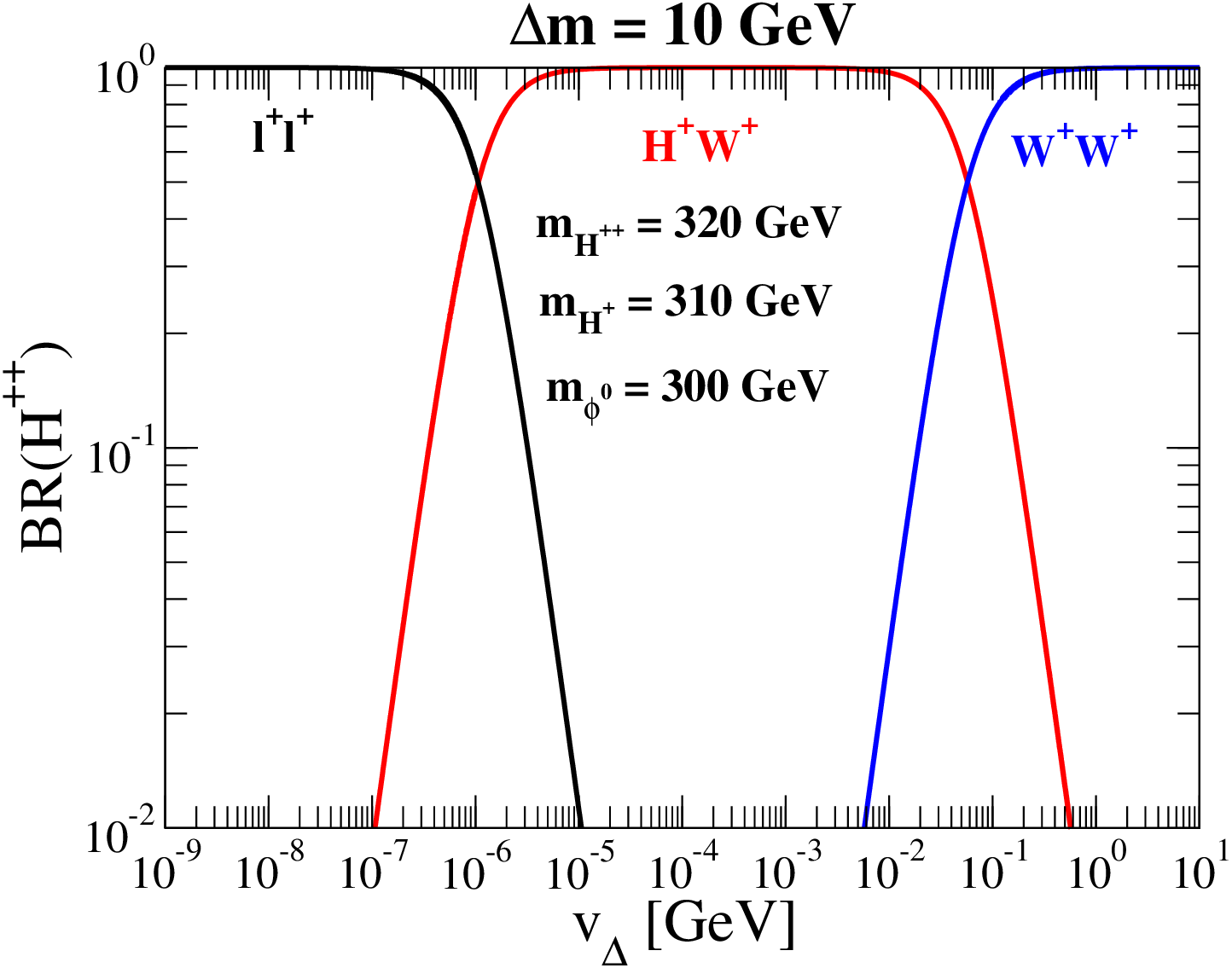}\\\vspace{7mm}
\includegraphics[width=65mm]{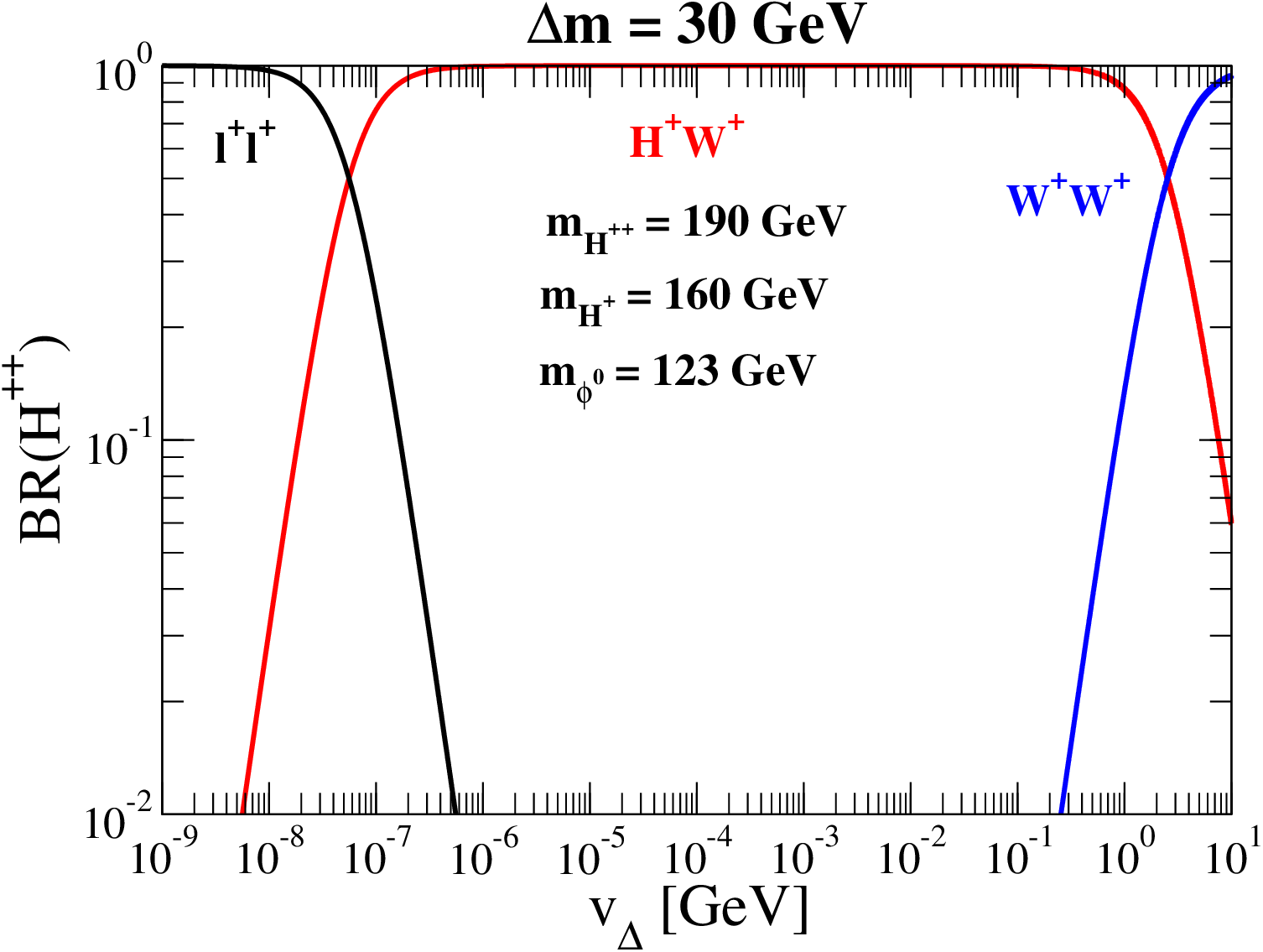}\hspace{3mm}
\includegraphics[width=65mm]{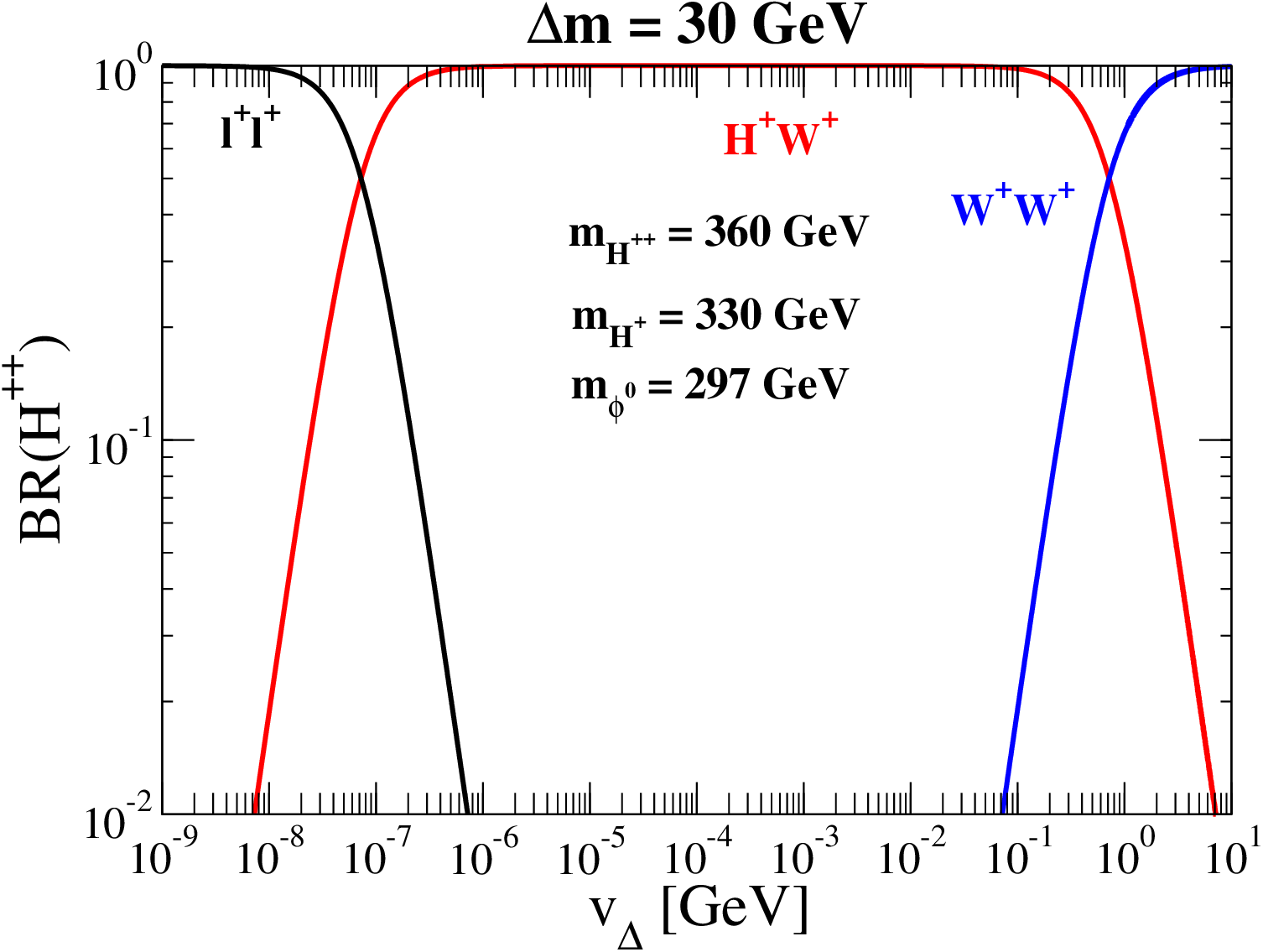}
\caption{Decay branching ratio of $H^{++}$ as a function of $v_\Delta$. 
In the upper left (right) figure, $m_{H^{++}}$ is 
fixed to be 120 GeV (300 GeV), and $\Delta m$ is taken to be zero. 
In the middle left (right) figure, $m_{H^{++}}$ is 
fixed to be 140 GeV (320 GeV), and $\Delta m$ is taken to be 10 GeV. 
In the bottom left (right) figure, 
$m_{H^{++}}$ is 
fixed to be 190 GeV (360 GeV), and $\Delta m$ is taken to be 30 GeV.}
\label{fig1}
\end{center}
\end{figure}
In this section, we discuss the decay branching ratios of the triplet-like scalar bosons $H^{\pm\pm}$, $H^\pm$, $H$ and $A$ \cite{Perez:2008ha}. 
We discuss both the cases of $\xi=0$ and $\xi>0$. 
At the end of this section, we also comment on the case of $\xi<0$. 
The decay modes of the triplet-like scalar bosons can be classified into three modes: 
(i) decay via the Yukawa coupling defined in Eq.~(\ref{yn}), 
(ii) that via $v_\Delta$ and 
(iii) that via the gauge coupling. 
The magnitude of the Yukawa coupling constant and $v_\Delta$ are related from the neutrino mass as in Eq.~(\ref{mn}). 
The main decay modes of $H^{++}$ and $H^+$ depend on the size of $v_\Delta$ and $\xi$. 
The decay mode (iii) particularly is important in the case of $\xi \neq 0$.  
Typically, in this case, the heaviest triplet-like scalar boson decays into the second heaviest one associated with the $W$ boson.  
The formulae of the decay rates of $H^{\pm\pm}$, $H^\pm$, $H$ and $A$ are listed in Appendix~A. 
Here, the leptonic decay modes through the Yukawa coupling $h_{ij}$ 
are summed over all flavors 
and each element of $h_{ij}$ is taken to be 0.1 eV/($\sqrt{2}v_\Delta$). 

In FIG.~\ref{decay-rate}, the decay width for the triplet-like scalar bosons is shown 
in the case of $\Delta m =$ 10 GeV and $\Delta m =$ 30~GeV. 
Since there is a decay mode through the gauge coupling 
the minimum value of the decay widths of $H^{++}$ and $H^+$ are $\mathcal{O}(10^{-6})$ GeV 
for $\Delta m=10$ GeV and $\mathcal{O}(10^{-4})$ GeV for $\Delta m=30$ GeV. 
On the other hand, the decay widths of $H$ and $A$ become minimum at $v_\Delta\simeq 10^{-4}-10^{-3}$ GeV 
with the magnitude of $\mathcal{O}(10^{-13}-10^{-12})$ GeV.  
This result is consistent with Ref.~\cite{Perez:2008ha}. 

We consider the decay branching ratio of $H^{++}$. 
In the case with $\Delta m =0$ and 
$m_{H^{++}}$=140 GeV, $H^{++}$ decays into $\ell^+\ell^+$ with $v_\Delta \lesssim 10^{-3}$ GeV 
or $W^+W^+$ with $v_\Delta \gtrsim 10^{-3}$ GeV. 
The value of $v_\Delta$ where the main decay mode changes from $H^{++}\to \ell^+\ell^+$ to $H^{++}\to W^+W^+$ is shifted 
at $v_\Delta \simeq 10^{-4}$ GeV when $m_{H^{++}}=300$ GeV. 
In the case of $\Delta m$ =10 GeV, 
$H^{++}$ decays into $H^+W^{+*}$ in the region of $10^{-6}\text{ GeV} \lesssim v_\Delta \lesssim 1\text{ GeV}$ 
($10^{-6}\text{ GeV} \lesssim v_\Delta \lesssim 0.1\text{ GeV}$) for $m_{H^{++}}$=140 GeV (320 GeV). 
Similarly, in the case of $\Delta m$ =30 GeV,  
$H^{++}$ decays into $H^+W^{+*}$ in the region of $10^{-7}\text{ GeV} \lesssim v_\Delta \lesssim 1\text{ GeV}$ 
for $m_{H^{++}}$=190 GeV and 360 GeV. 
In FIG.~\ref{fig1}, the decay branching ratio of $H^{++}$ is shown as 
a function of $v_\Delta$. 

\begin{figure}[t]
\begin{center}
\includegraphics[width=65mm]{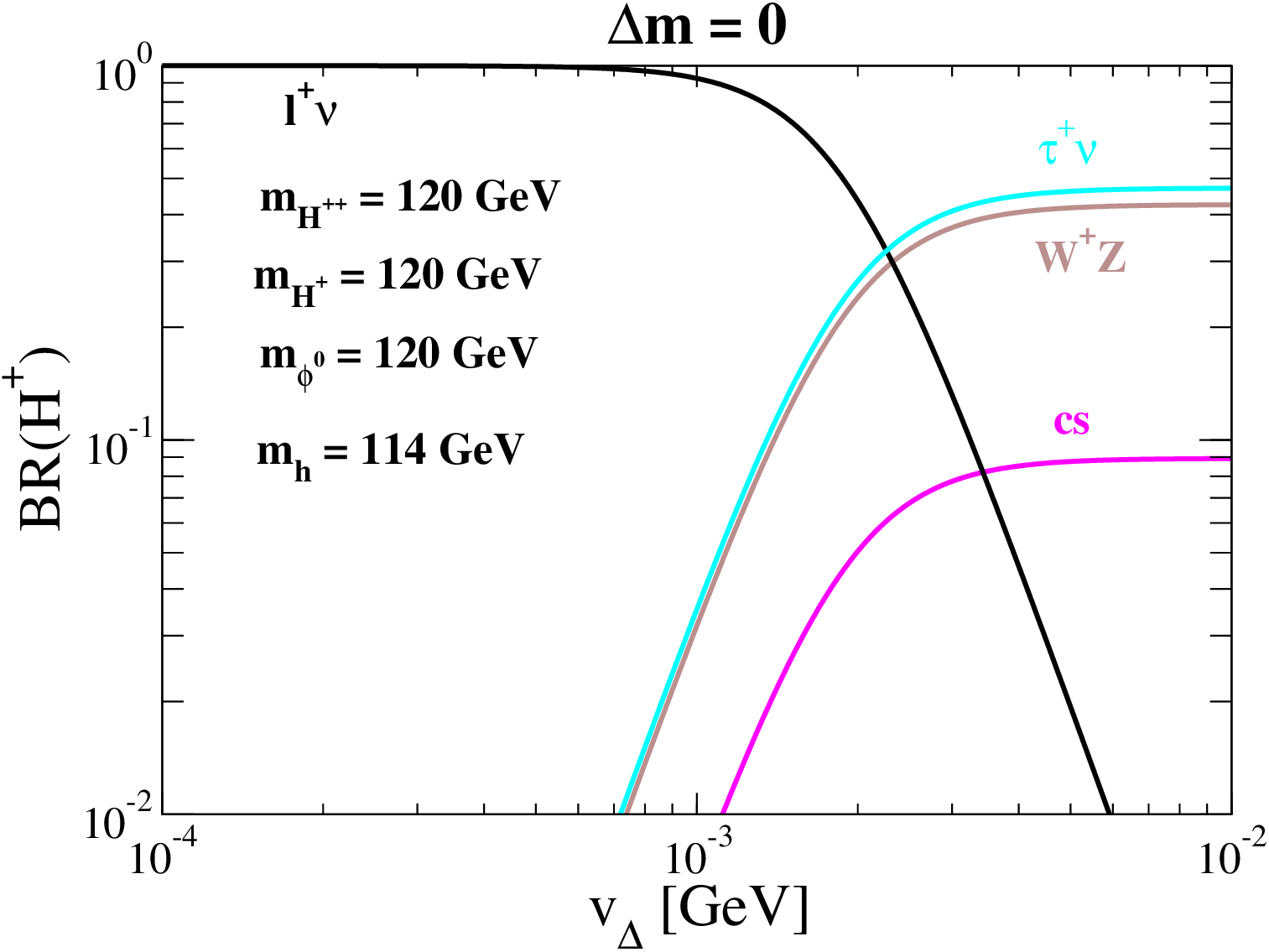}\hspace{3mm}
\includegraphics[width=65mm]{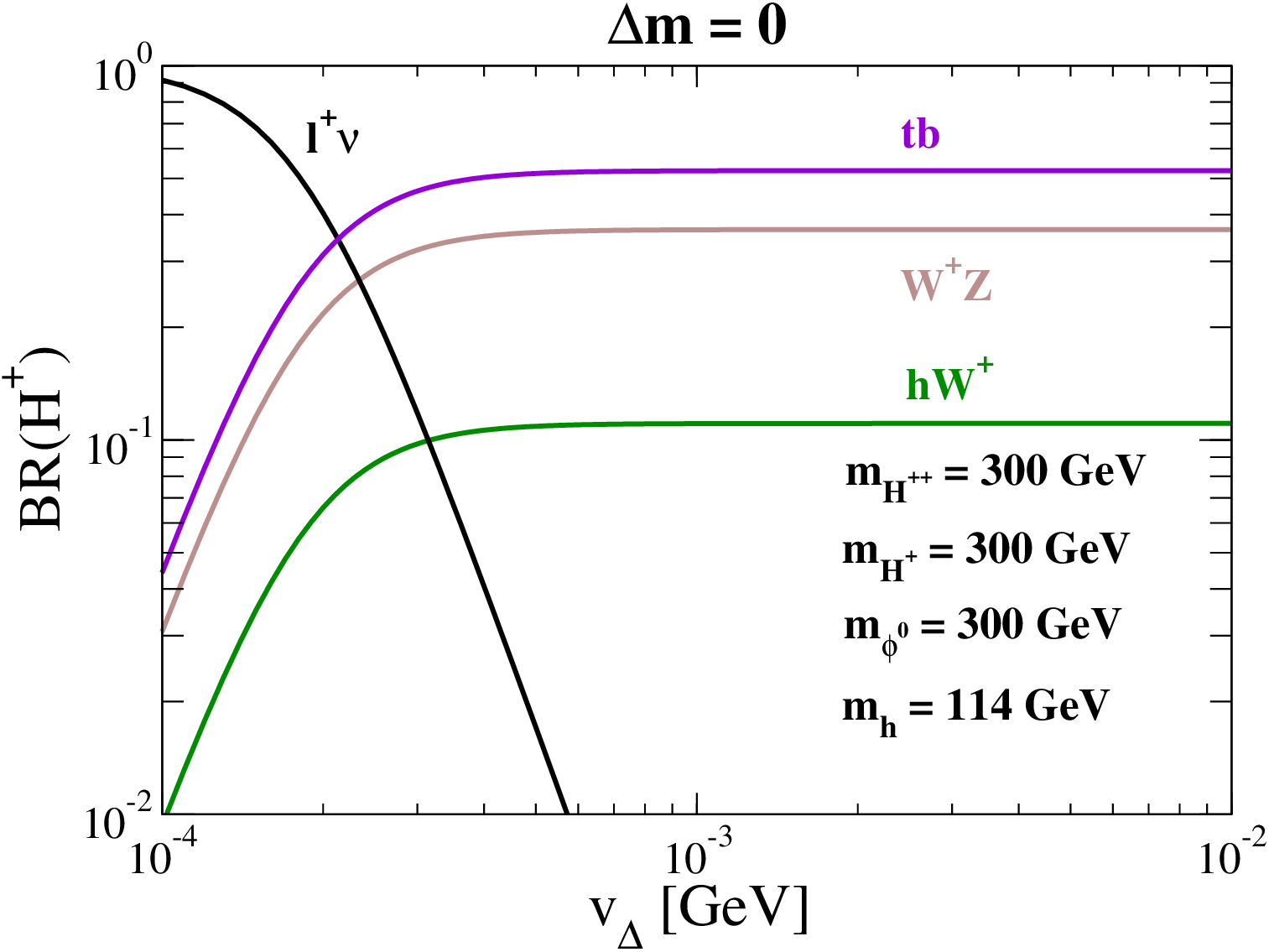}\\
\vspace{7mm}
\includegraphics[width=65mm]{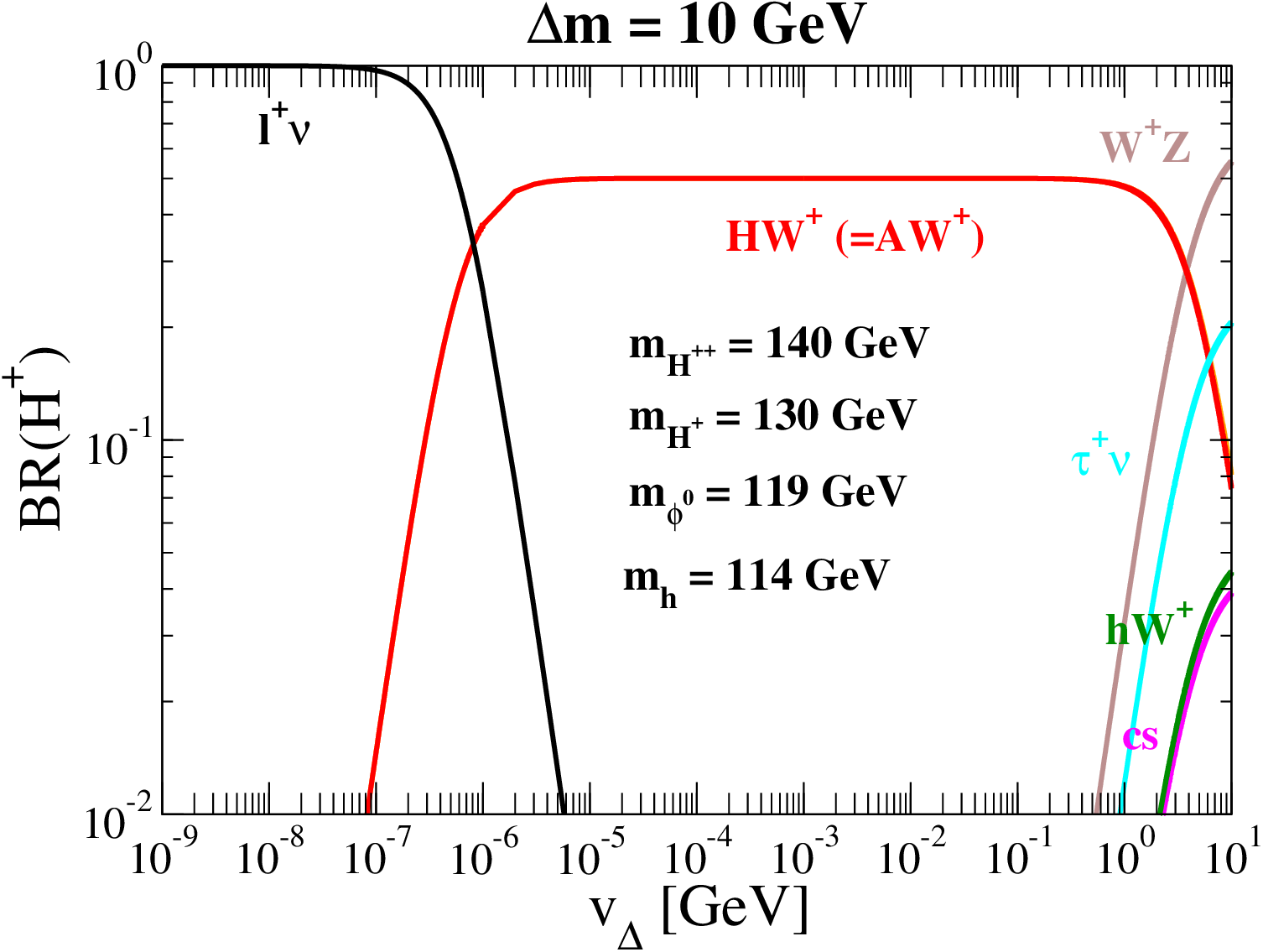}\hspace{3mm}
\includegraphics[width=65mm]{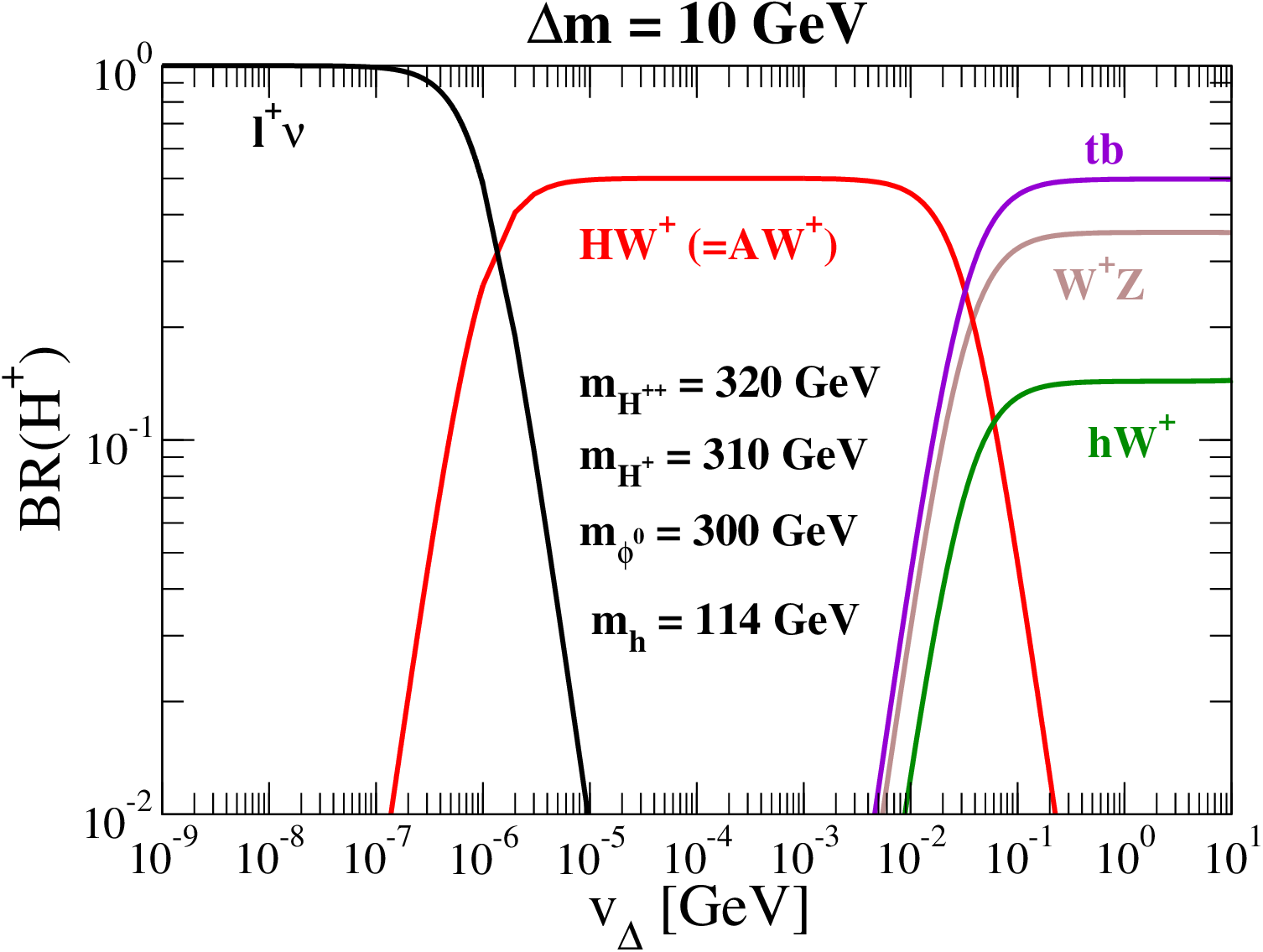}\\
\vspace{7mm}
\includegraphics[width=65mm]{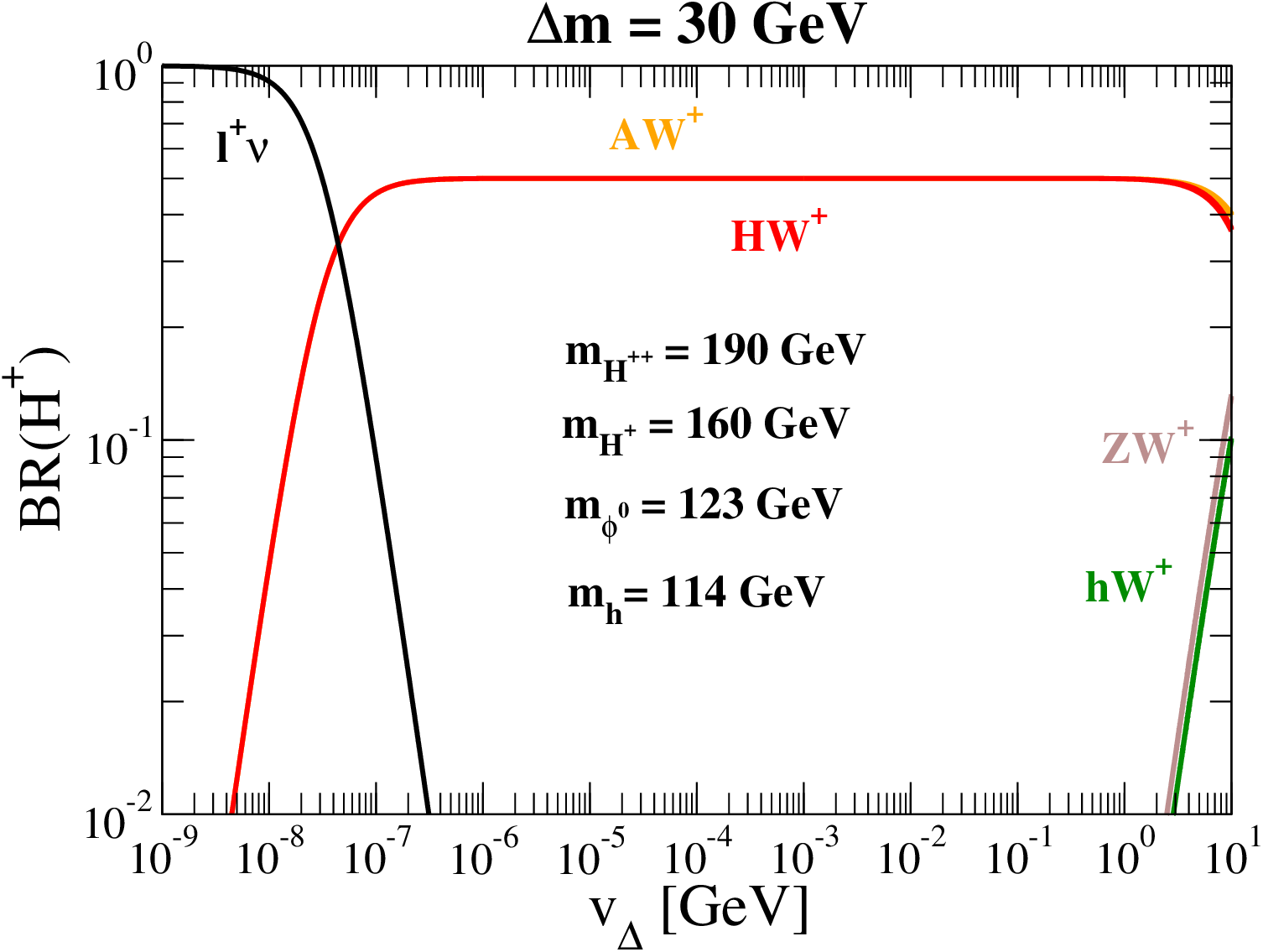}\hspace{3mm}
\includegraphics[width=65mm]{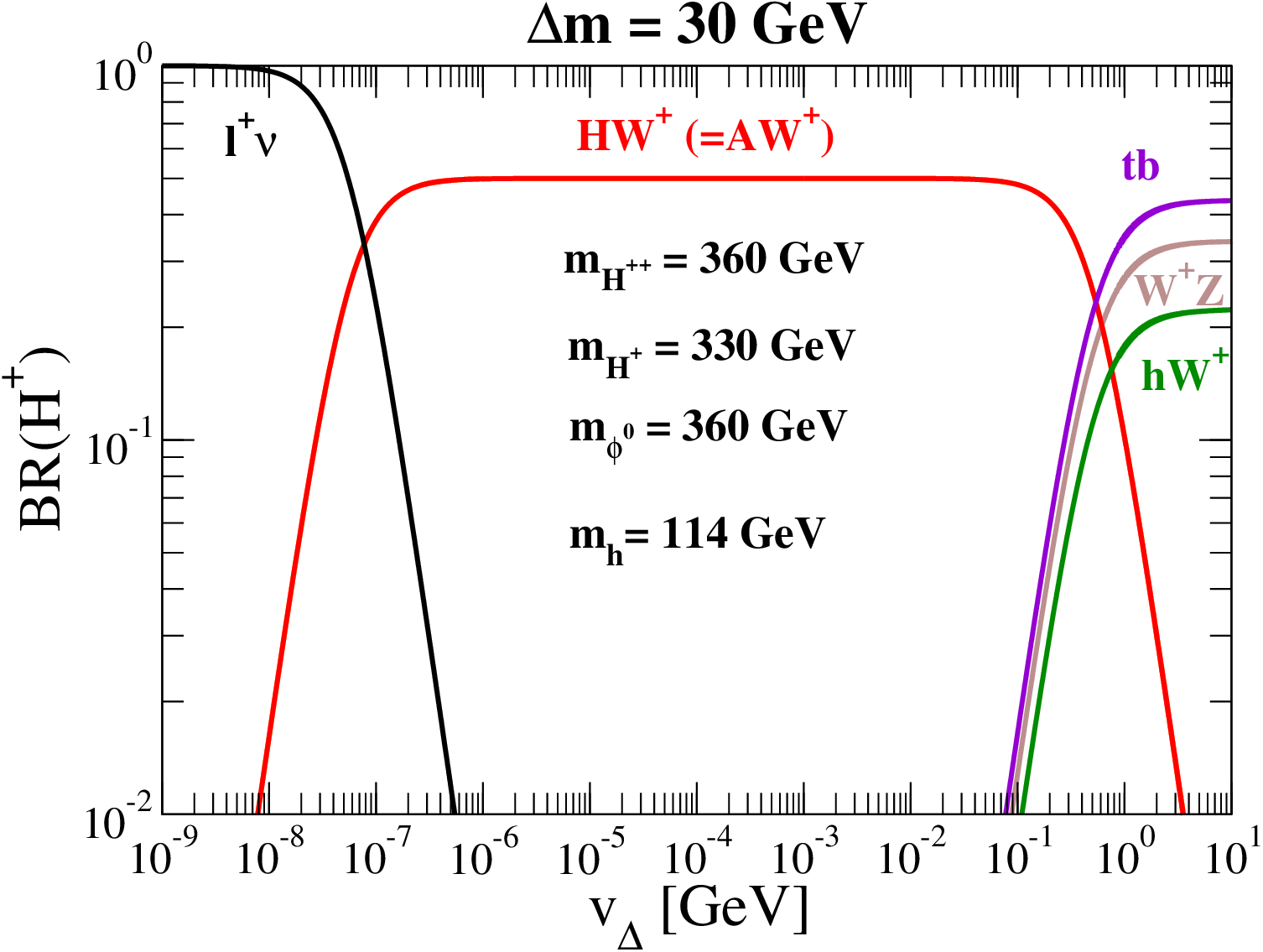}
\caption{Decay branching ratio of $H^+$ as a function of $v_\Delta$. 
In all the figures, $m_h$ is taken to be 114 GeV. 
In the upper left (right) figure, $m_{H^+}$ is 
fixed to be 120 GeV (300 GeV), and $\Delta m$ is taken to be zero. 
In the middle left (right) figure, $m_{H^+}$ is 
fixed to be 130 GeV (310 GeV), and $\Delta m$ is taken to be 10 GeV. 
In the bottom left (right) figure, $m_{H^+}$ is 
fixed to be 160 GeV (330 GeV), and $\Delta m$ is taken to be 30 GeV.}
\label{fig2}
\end{center}
\end{figure}

The decay branching ratio of $H^+$ is shown in FIG.~\ref{fig2}. 
In the case of $\Delta m=0$, $H^+$ decays into $\ell^+\nu$ with $v_\Delta < 10^{-4}-10^{-3}$ GeV.  
When $v_\Delta > 10^{-4}-10^{-3}$ GeV, 
$H^+$ decays into $\tau^+\nu$, $W^+ Z$ and $c\bar{s}$ for $m_{H^+}=120$ GeV, while 
$H^+$ decays into $t\bar{b}$, $W^+ Z$ and $hW^+$ for $m_{H^+}=300$ GeV. 
In the case of $\Delta m$ =10 GeV, similarly to the decay of $H^{++}$, $H^+$ decays into $\phi^0 W^{+*} $ 
in the region of $10^{-6}\text{ GeV} \lesssim v_\Delta \lesssim 1\text{ GeV}$ 
($10^{-6}\text{ GeV} \lesssim v_\Delta \lesssim 10^{-2}\text{ GeV}$ )
for $m_{H^+}=130$ GeV (310 GeV). 
In the case of $\Delta m$ =30 GeV, $H^+$ decays into $\phi^0 W^{+*} $ 
in the region of $10^{-7}\text{ GeV} \lesssim v_\Delta \lesssim 10\text{ GeV}$ 
($10^{-7}\text{ GeV} \lesssim v_\Delta \lesssim 10^{-1}\text{ GeV}$) for $m_{H^+}=160$ GeV (330 GeV).

\begin{figure}[t]
\begin{center}
\includegraphics[width=65mm]{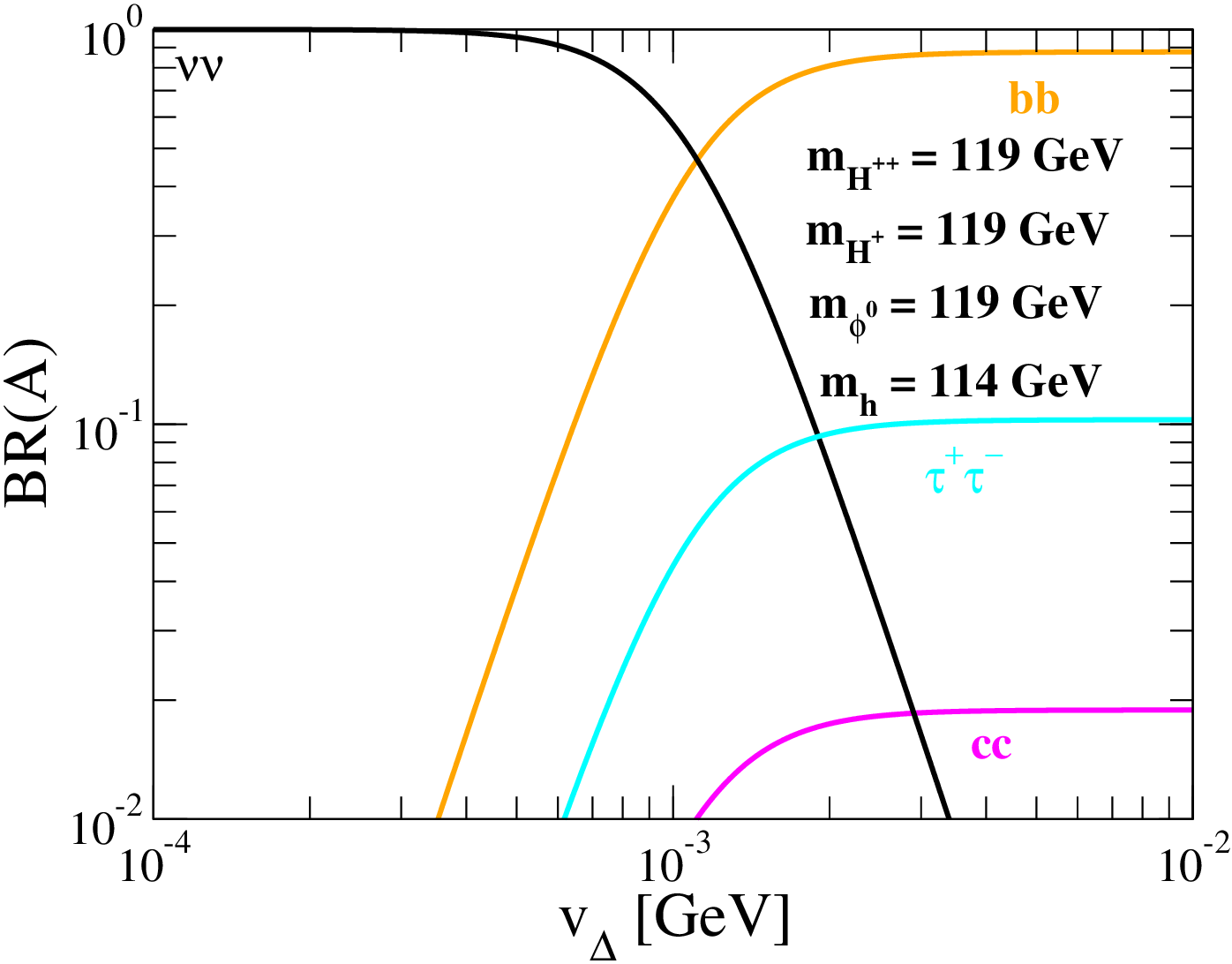}\hspace{3mm}
\includegraphics[width=66mm]{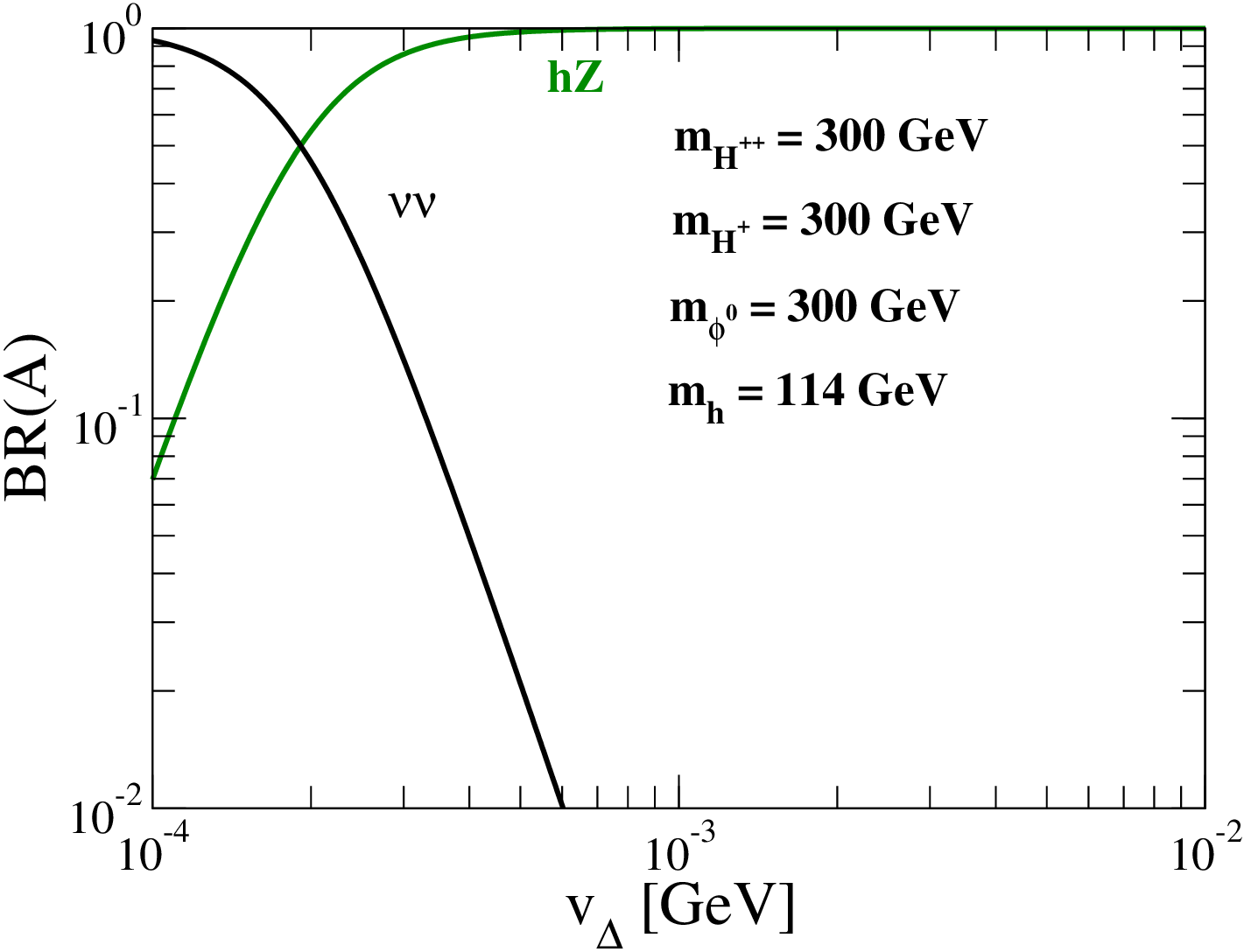}\\
\vspace{7mm}
\includegraphics[width=67mm]{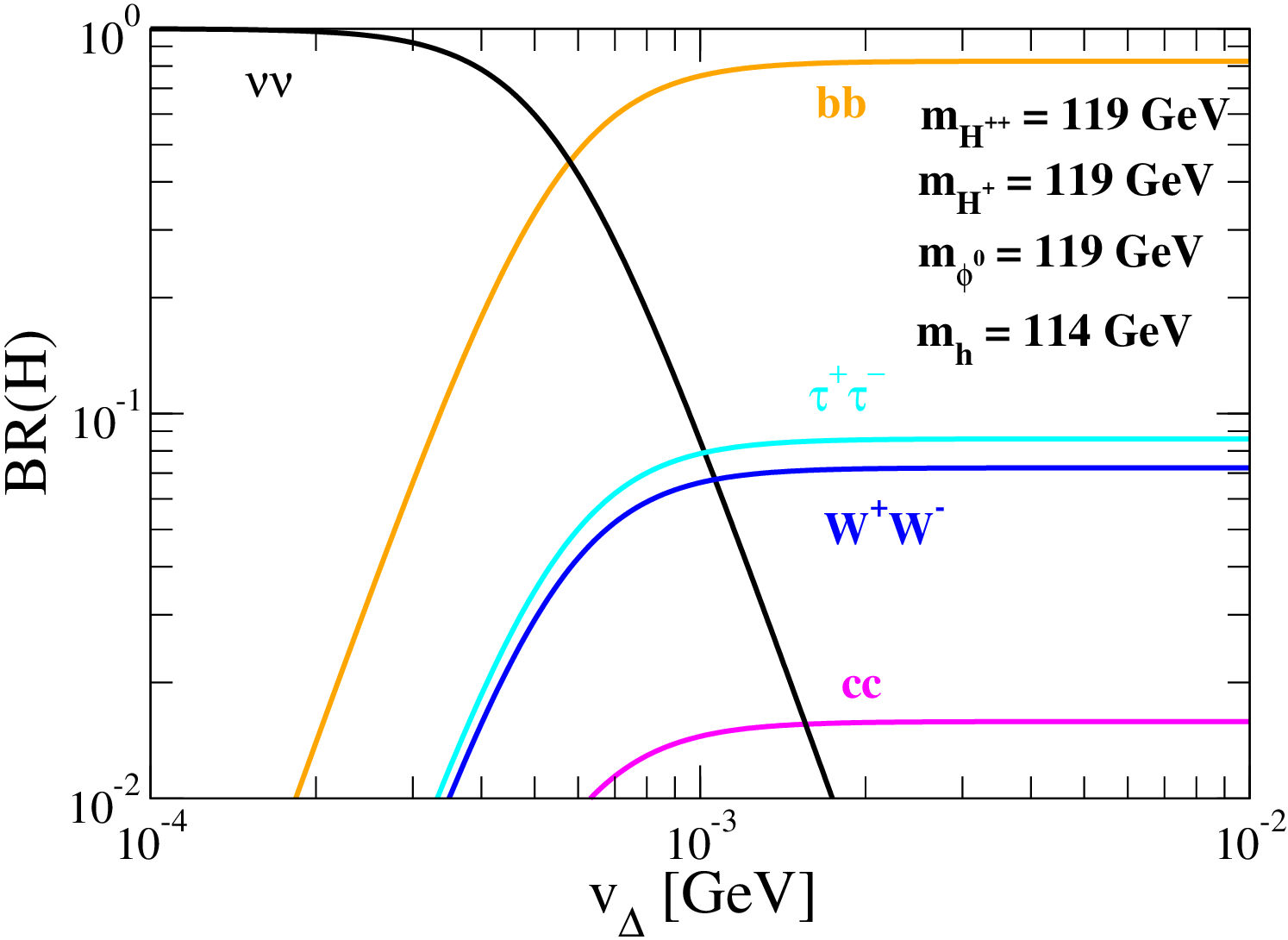}\hspace{3mm}
\includegraphics[width=65mm]{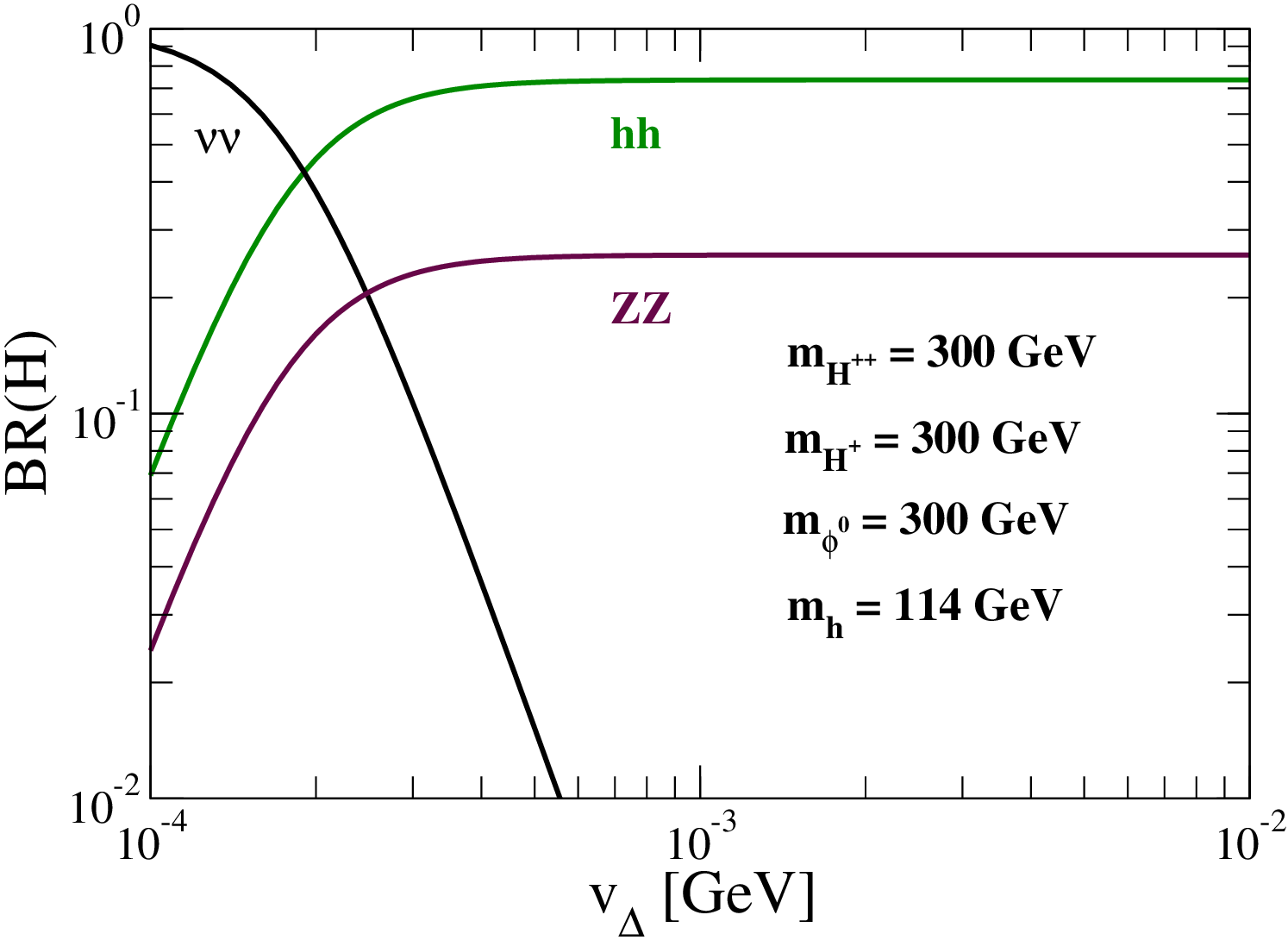}\\
\caption{Decay branching ratios of $A$ and $H$ as a function of $v_\Delta$. 
In all the figures, $m_h$ is taken to be 114 GeV. 
In the upper left (right) figure, the branching ratio of $A$ is shown in the case of $m_A=119$ GeV (300 GeV). 
In the lower left (right) figure, the branching ratio of $H$ is shown in the case of $m_H=119$ GeV (300 GeV). }
\label{fig3}
\end{center}
\end{figure}

The decay branching ratios of $H$ and $A$ are shown in FIG.~\ref{fig3}. 
Both $H$ and $A$ decay into neutrinos in the region of $v_\Delta < 10^{-4}-10^{-3}$ GeV. 
When $v_\Delta> 10^{-4}-10^{-3}$ GeV, 
both $H$ and $A$ decay into $b\bar{b}$ with $m_{\phi^0} = 119$ GeV 
while $H$ ($A$) decay into $hh$ and $ZZ$ ($hZ$) with $m_{\phi^0} = 300$ GeV. 

Finaly, we comment on the case of $\xi <0$. 
In this case, $H$ and $A$ can decay into $H^\pm W^{\mp(*)} $ depending on the magnitude of 
$\xi$ and $v_\Delta$. 
At the same time, 
$H^+$ can decay into $H^{++} W^{-(*)}$. 
The decay of $H^{++}$ is the 
same as in the case without the mass difference. 

\section{Mass determination of the triplet-like scalar bosons at the LHC}

\begin{figure}[t]
\begin{center}
\includegraphics[width=57mm]{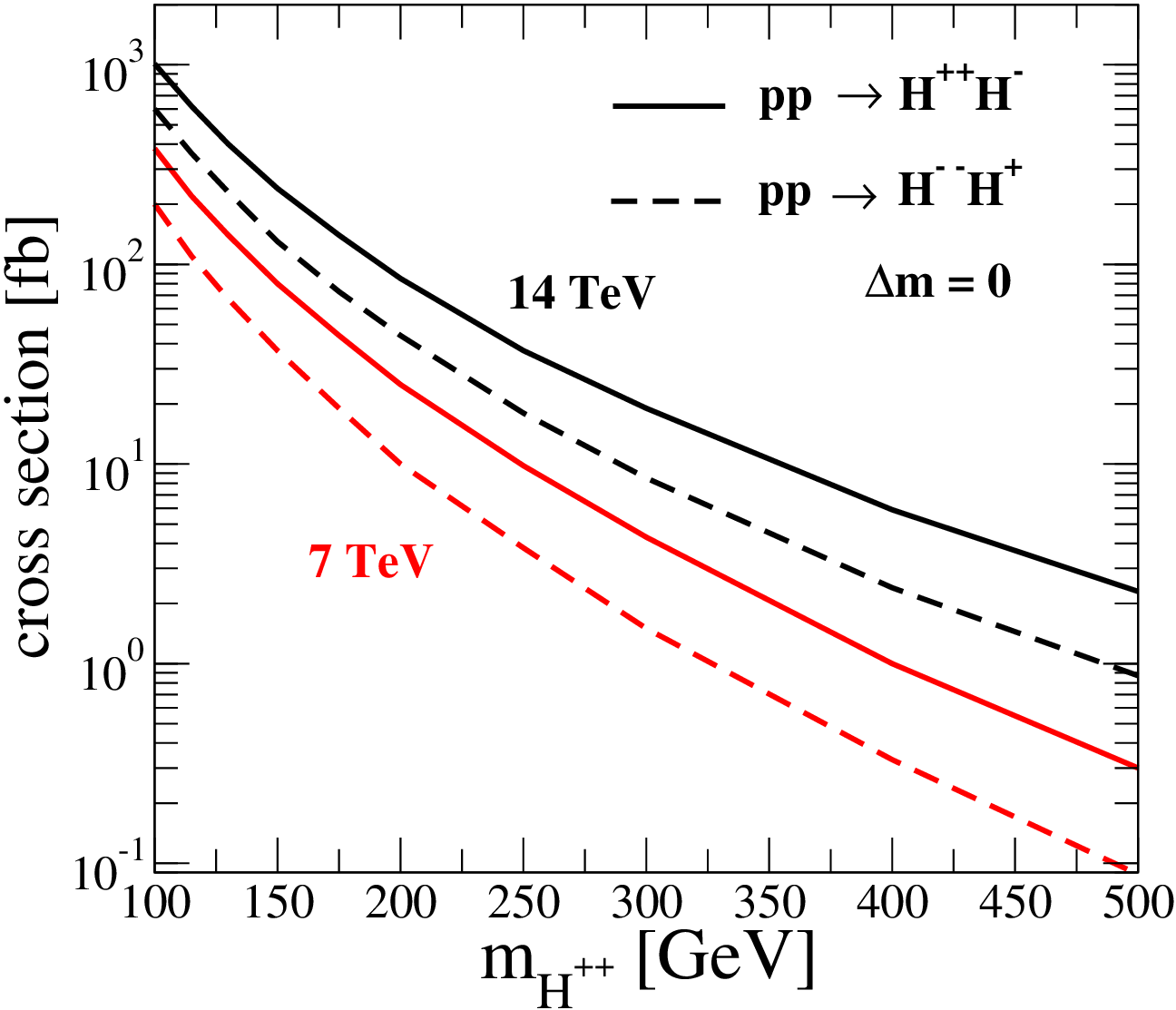}\hspace{2mm}
\includegraphics[width=57mm]{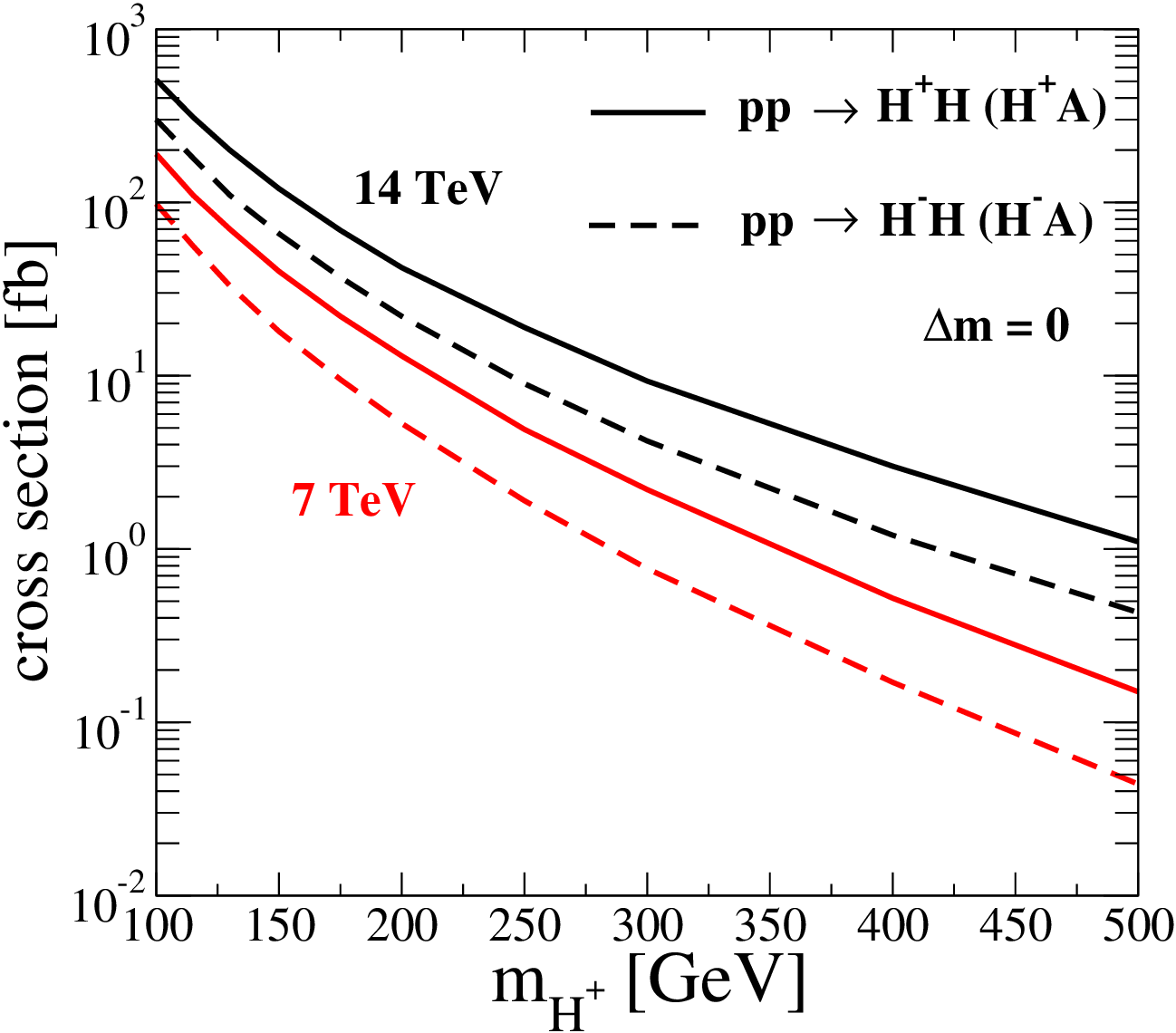}\hspace{2mm}
\includegraphics[width=57mm]{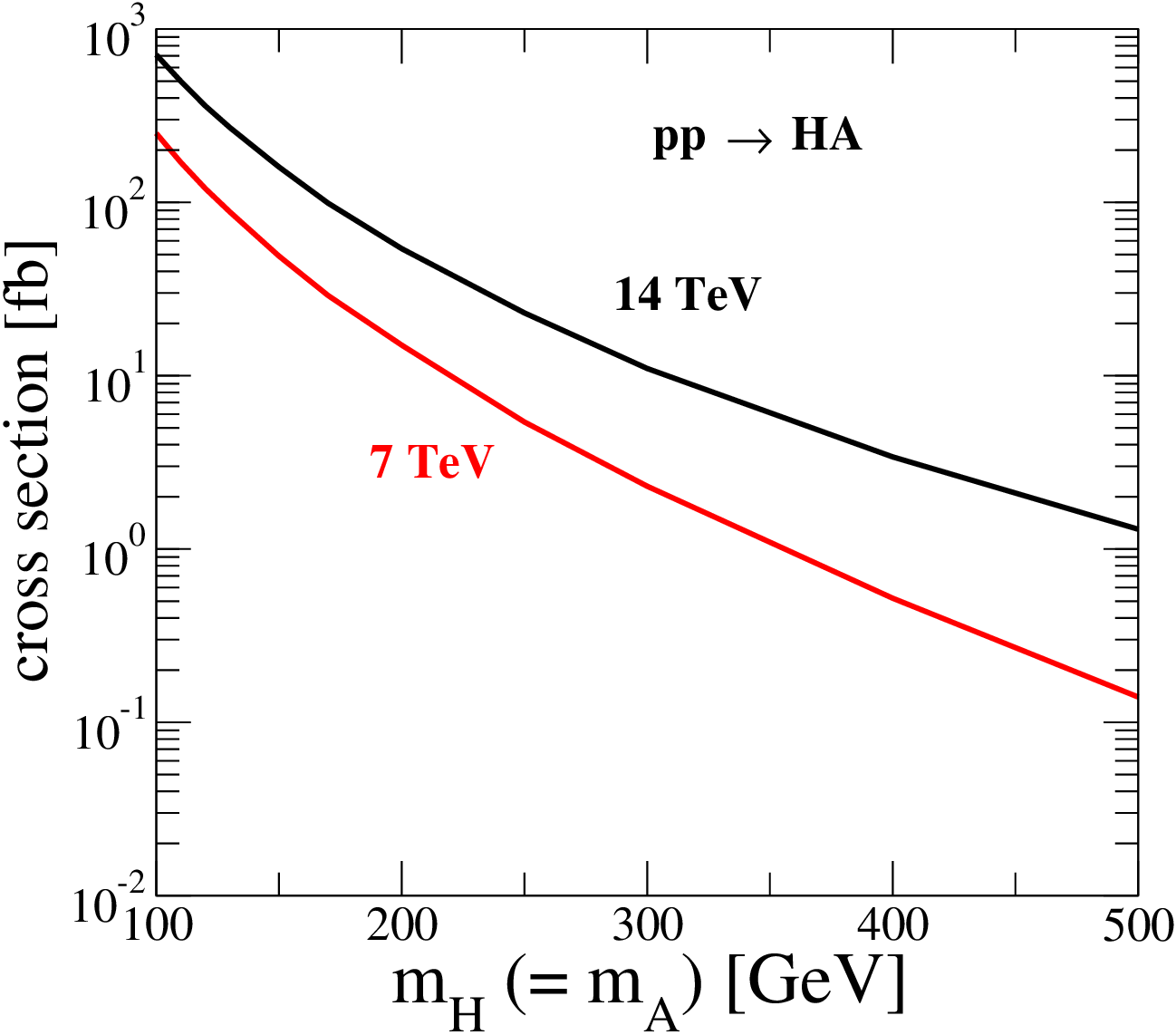}
\caption{Production cross sections for the triplet-like scalar bosons in the Drell-Yan process. }
\label{cs}
\end{center}
\end{figure}

In this section, we discuss how the HTM with $\xi > 0$ can be tested at the LHC.  
At the LHC, the triplet-like scalar bosons $H^{\pm\pm}$, $H^\pm$, $H$ and $A$ 
are mainly produced through 
the Drell-Yan processes, for instance, $pp\to H^{++}H^{--}$, $pp\to H^+H^-$, 
$pp\to H^{\pm\pm}H^\mp$ and $pp\to H^\pm \phi^0$ and $pp\to HA$. 
In particular, latter three processes are important when we consider the case of $\xi >0$. 
The cross sections for the latter three production processes are shown in FIG.~\ref{cs}. 

We comment on vector boson fusion production processes. 
There are two types of the vector boson fusion processes. 
First one is the process via $VV\Delta$ vertices, where $V=Z$ or $W^\pm$. 
The cross section of this process is small, since the $VV\Delta$ vertex is proportional to $v_\Delta$
\footnote{The magnitude of $v_\Delta$ may be determined indirectly via $B_{ee}/B_{WW}$ or 
$\Gamma_{ee}$ and $0\nu\beta\beta$ where $H^{++}\to \ell^+\ell^+$, $W^+W^+$ are dominant \cite{Kadastik:2007yd}. 
On the other hand, it could be 
directly measured via $qq\to q'q W^{\pm*}Z^*\to q'q H^\pm$
at the LHC~\cite{Asakawa:2005gv} and via $e^+e^-\to Z^* \to H^\pm W^\mp$ at the ILC~\cite{Cheung:1994rp}.}. 
The other one is the process via the gauge coupling constant. 
In particular, 
$qq\to q'q'H^{++}\phi^0$ is the unique process 
whose difference of the electric charge between produced scalar bosons is two.  
This production cross section is 0.51 fb (0.13 fb) at $\sqrt{s}=14$ TeV ($\sqrt{s}=7$ TeV ) assuming 
mass parameters Set~1 which is given just below.  

We consider the following two sets for mass parameters: 
\begin{align}
\text{(Set~1)}\quad
m_{H^{++}}=140 \text{ GeV},\quad m_{H^+}=130 \text{ GeV},\quad m_H=m_A=119 \text{ GeV},\quad m_h = 114 \text{ GeV},\notag\\
\text{(Set~2)}\quad
m_{H^{++}}=190 \text{ GeV},\quad m_{H^+}=160 \text{ GeV},\quad m_H=m_A=123 \text{ GeV},\quad m_h = 114 \text{ GeV},\notag
\end{align}
which correspond to the cases with $\xi$=(52 GeV)$^2$ and $\xi$=(102 GeV)$^2$, respectively. 
In the following numerical analysis, $\lambda_2=0$ is taken.  
In these parameter sets, the production cross sections for the triplet-like scalar bosons are listed in TABLE~\ref{cs1}.  
We can classify scenarios by the following four regions of $v_\Delta$ for Set~1: 
\begin{align}
\begin{array}{cl}
\text{Scenario (1a)} &\quad v_\Delta \gtrsim 1 \text{ GeV} ,\\
\text{Scenario (1b)} &\quad 10^{-3} \text{ GeV} \lesssim v_\Delta \lesssim 1 \text{ GeV}, \\
\text{Scenario (1c)} &\quad 10^{-6} \text{ GeV} \lesssim v_\Delta \lesssim 10^{-3} \text{ GeV}, \\
\text{Scenario (1d)} &\quad v_\Delta \lesssim 10^{-6} \text{ GeV}. 
\end{array}\notag
\end{align}
We can also classify scenarios by the following four regions of $v_\Delta$ for Set~2: 
\begin{align}
\begin{array}{cl}
\text{Scenario (2a)} &\quad v_\Delta \gtrsim 1 \text{ GeV} ,\\
\text{Scenario (2b)} &\quad 10^{-4} \text{ GeV} \lesssim v_\Delta \lesssim 1 \text{ GeV}, \\
\text{Scenario (2c)} &\quad 10^{-7} \text{ GeV} \lesssim v_\Delta \lesssim 10^{-4} \text{ GeV}, \\
\text{Scenario (2d)} &\quad v_\Delta \lesssim 10^{-7} \text{ GeV}. 
\end{array}\notag
\end{align}

\begin{table}[t]
\begin{center}
{\renewcommand\arraystretch{1.2}
\begin{tabular}{|l||c|c|c|}\hline
Process &$\Delta m=0$ at $\sqrt{s}=14$ TeV (7 TeV)&$\Delta m=$10 GeV at $\sqrt{s}=14$ TeV (7 TeV)
& $\Delta m=$30 GeV  at $\sqrt{s}=14$ TeV (7 TeV) 
\\\hline\hline
$pp\to  H^{++}H^-$&310 fb (110 fb)&350 fb (120 fb)&140 fb (43 fb)\\\hline
$pp\to  H^+H$     &150 fb (53 fb) &230 fb (81 fb) &150 fb (50 fb)\\\hline
$pp\to  HA  $     &200 fb (65 fb) 
&370 fb (130 fb) 
&330 fb (110 fb)\\\hline
\end{tabular}}
\caption{Production cross sections for the triplet-like scalar bosons 
in the case of $\Delta m =0$ with $m_{H^{++}}=140$ GeV, 
those of the case for Set~1 and Set~2. }
\label{cs1}
\end{center}
\end{table}

\begin{table}[t]
\begin{center}
{\renewcommand\arraystretch{1.2}
\begin{tabular}{|c||c|c|c|c|}\hline
Scenario &Decay of $H^{++}$&Decay of $H^+$&Decay of $H$&Decay of $A$ \\\hline\hline
\hspace{-5.5mm}(1a) ($v_\Delta = 5$ GeV)&$W^+W^{+*}$ [0.93]&$W^{+*}Z$ [0.37], $\tau^+\nu$ [0.14] &$b\bar{b}$ [0.82]&$b\bar{b}$ [0.89]\\\hline
(1b) ($v_\Delta = 10^{-2}$ GeV)&$H^+W^{+*}$ [1.0]&$AW^{+*}$ [0.5], $HW^{+*}$ [0.5]&$b\bar{b}$ [0.82]&$b\bar{b}$ [0.89]\\\hline
(1c) ($v_\Delta = 10^{-5}$ GeV)&$H^+W^{+*}$ [1.0]&$AW^{+*}$ [0.5], $HW^{+*}$ [0.5]&$\nu\nu$ [1.0]&$\nu\nu$ [1.0]\\\hline
(1d) ($v_\Delta = 10^{-8}$ GeV)&$\ell^+\ell^+$ [1.0]&$\ell^+\nu$ [1.0]&$\nu\nu$ [1.0]&$\nu\nu$ [1.0] \\\hline
\end{tabular}}
\caption{The main decay mode of the triplet-like scalar bosons in Scenario (1a) to Scenario (1d). 
The masses of the triplet-like scalar bosons are taken to be as for Set~1. 
The number in ( ) represents the sample value of $v_\Delta$ corresponding to the scenario. 
The number in [ ] represents the value of the decay branching ratio corresponding to the 
value of $v_\Delta$ displayed in ( ) in the same row. 
Here, $\ell\ell$ mode and $\ell\nu$ mode are summed over all flavors. }
\label{tdecay1}
\end{center}
\end{table}
\begin{table}[t]
\begin{center}
{\renewcommand\arraystretch{1.2}
\begin{tabular}{|l||c|c|c|c|}\hline
Scenario &Decay of $H^{++}$&Decay of $H^+$&Decay of $H$&Decay of $A$ \\\hline\hline
(2a) [$v_\Delta = 5$ GeV]&$W^+W^{+*}$ [0.76]&$AW^{+*}$ [0.47] $HW^{+*}$ [0.46] &$b\bar{b}$ [0.78]&$b\bar{b}$ [0.89]\\\hline
(2b) [$v_\Delta = 10^{-2}$ GeV]&$H^+W^{+*}$ [1.0]&$AW^{+*}$ [0.5] $HW^{+*}$ [0.5]&$b\bar{b}$ [0.78]&$b\bar{b}$ [0.89]\\\hline
(2c) [$v_\Delta = 10^{-5}$ GeV]&$H^+W^{+*}$ [1.0]&$AW^{+*}$ [0.5] $HW^{+*}$ [0.5]&$\nu\nu$ [1.0]&$\nu\nu$ [1.0]\\\hline
(2d) [$v_\Delta = 10^{-8}$ GeV]&$\ell^+\ell^+$ [0.97]&$\ell^+\nu$ [0.91]&$\nu\nu$ [1.0]&$\nu\nu$ [1.0] \\\hline
\end{tabular}}
\caption{
The main decay mode of the triplet-like scalar bosons in Scenario (2a) to Scenario (2d). 
The masses of the triplet-like scalar bosons are taken to be as for Set~2. 
The number in ( ) represents the sample value of $v_\Delta$ corresponding to the scenario. 
The number in [ ] represents the value of the decay branching ratio corresponding to the 
value of $v_\Delta$ displayed in ( ) in the same row. 
Here, $\ell\ell$ mode and $\ell\nu$ mode are summed over all flavors. }
\label{tdecay2}
\end{center}
\end{table}
\noindent
In each scenario, main decay modes of the triplet-like scalar bosons are listed in TABLE~\ref{tdecay1} 
and TABLE~\ref{tdecay2}. 
We here analyse the signal for Set~1 which may be used to reconstruct the masses of the triplet-like scalar bosons. 
The signal distributions discussed below are calculated by using CalcHEP~\cite{Pukhov:2004ca}. 

\begin{figure}[t]
\begin{center}
\includegraphics[width=65mm]{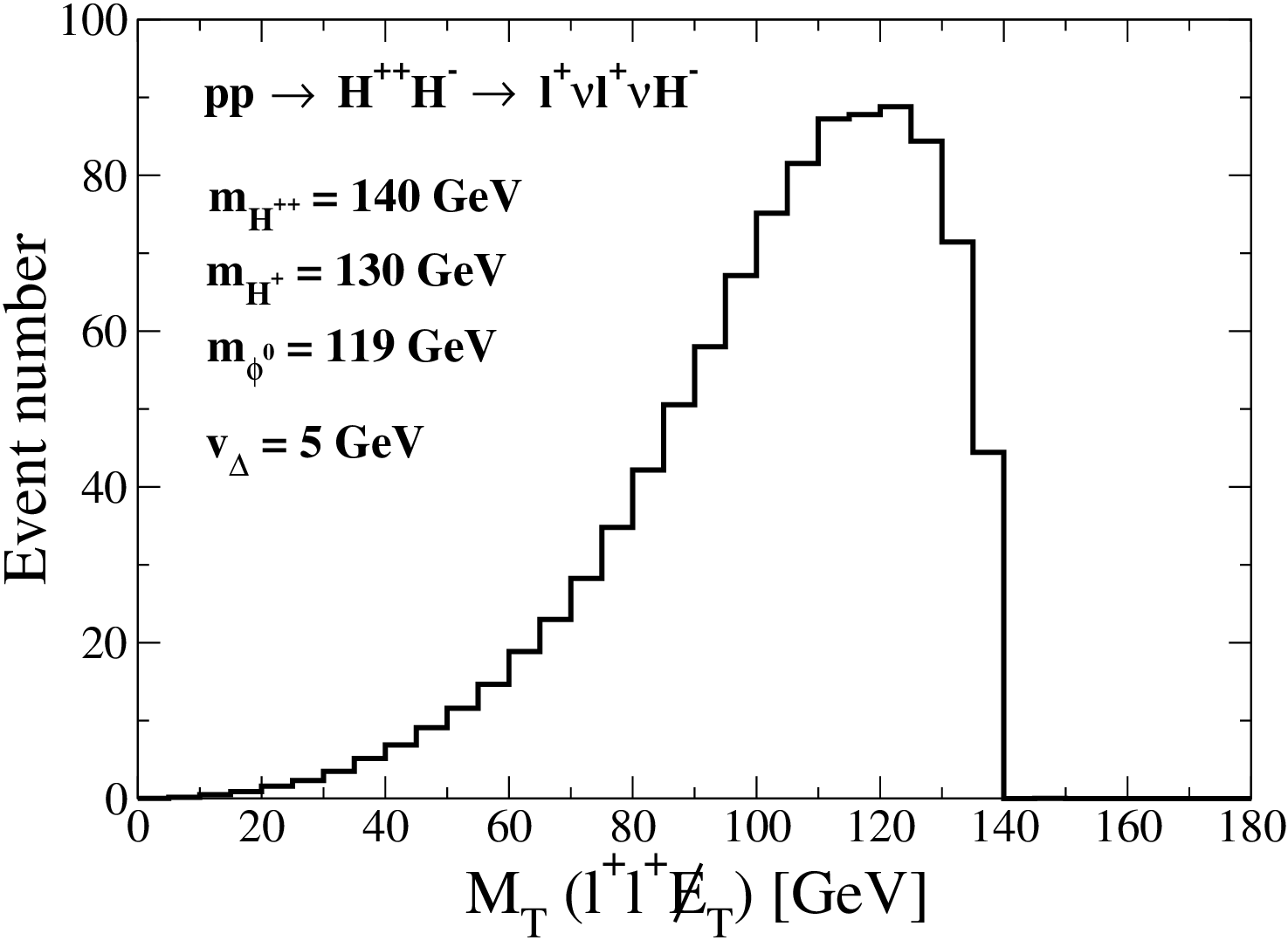}\hspace{3mm}
\includegraphics[width=65mm]{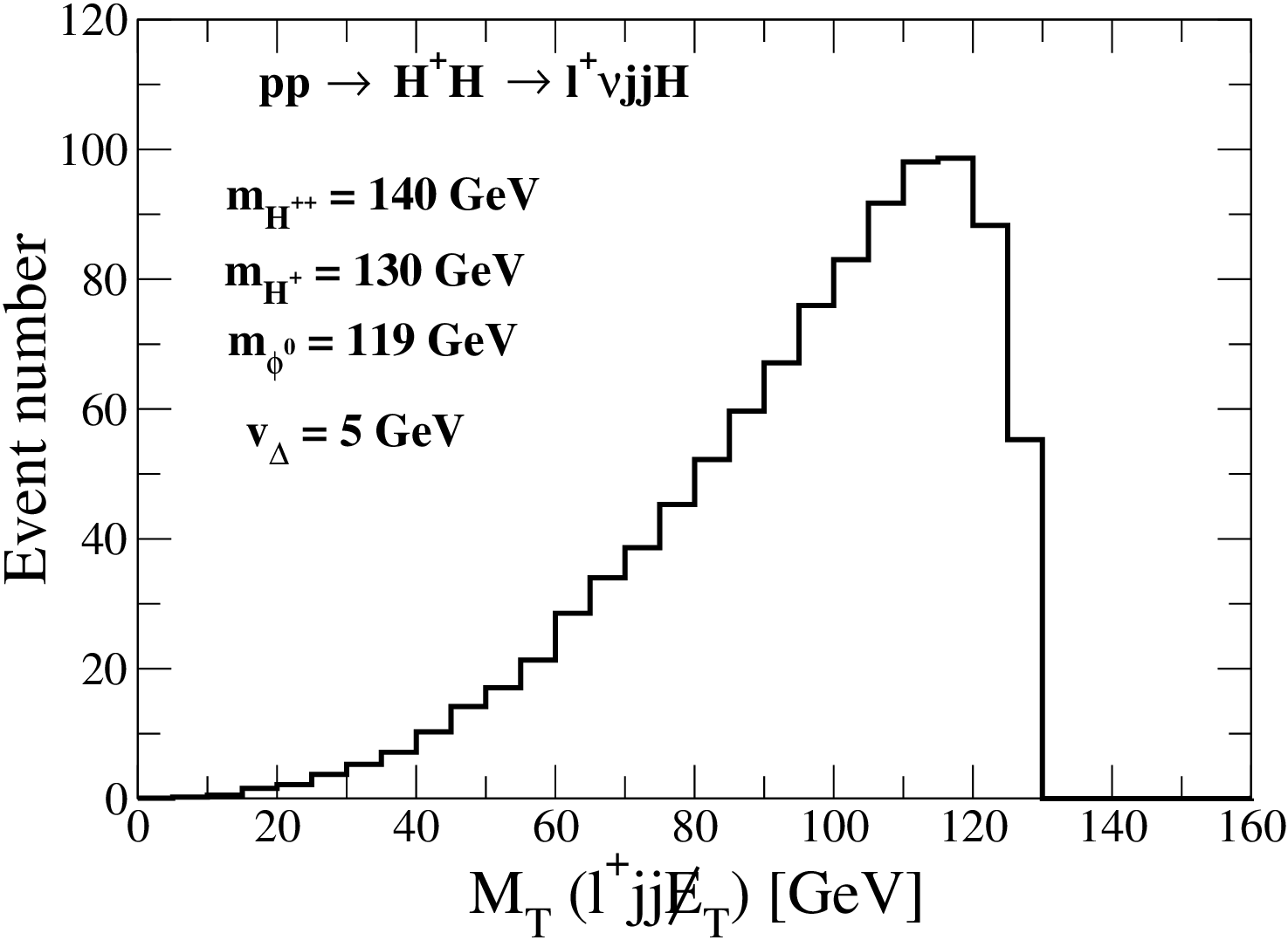}\\\vspace{7mm}
\includegraphics[width=65mm]{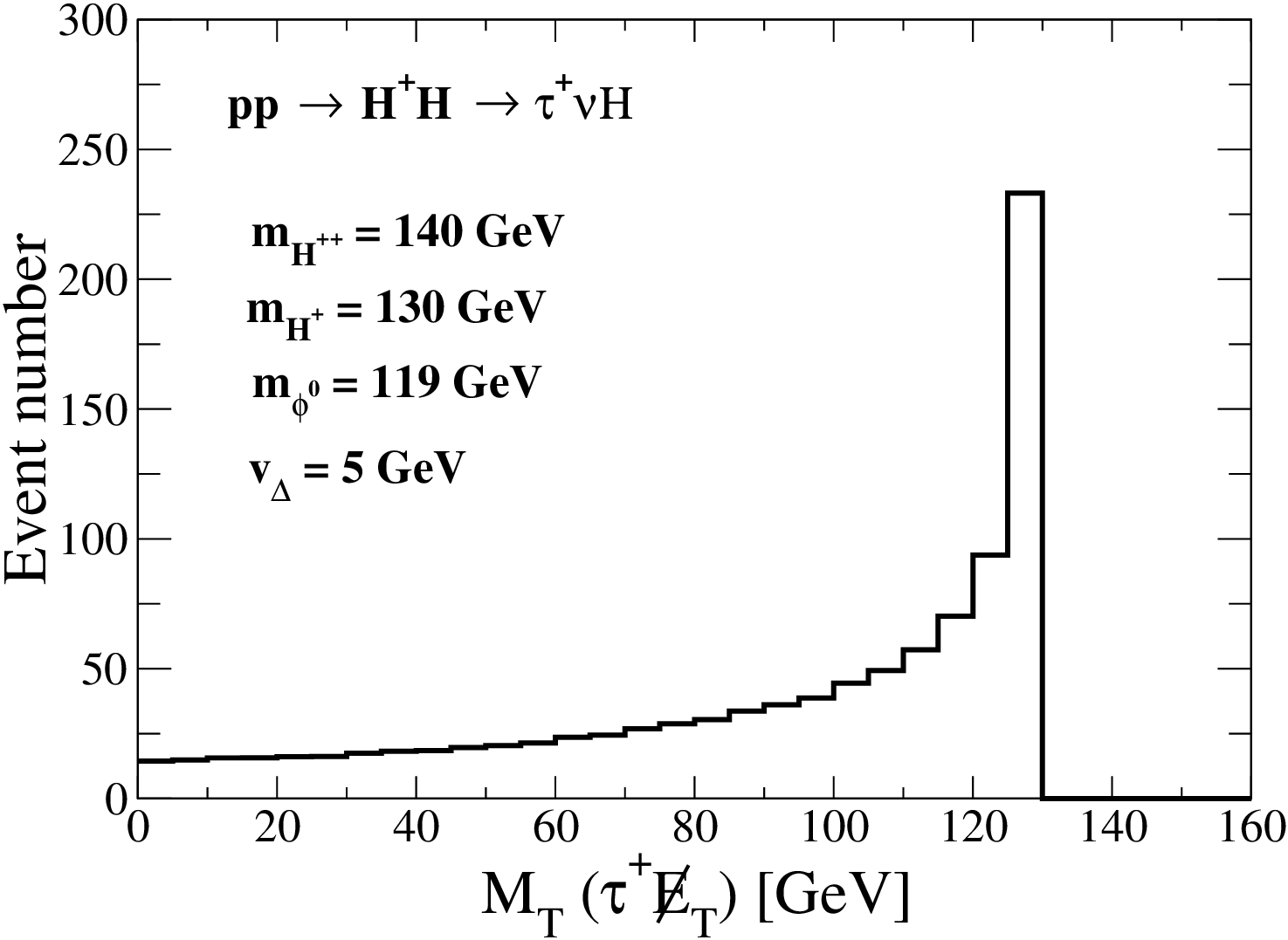}\hspace{3mm}
\includegraphics[width=65mm]{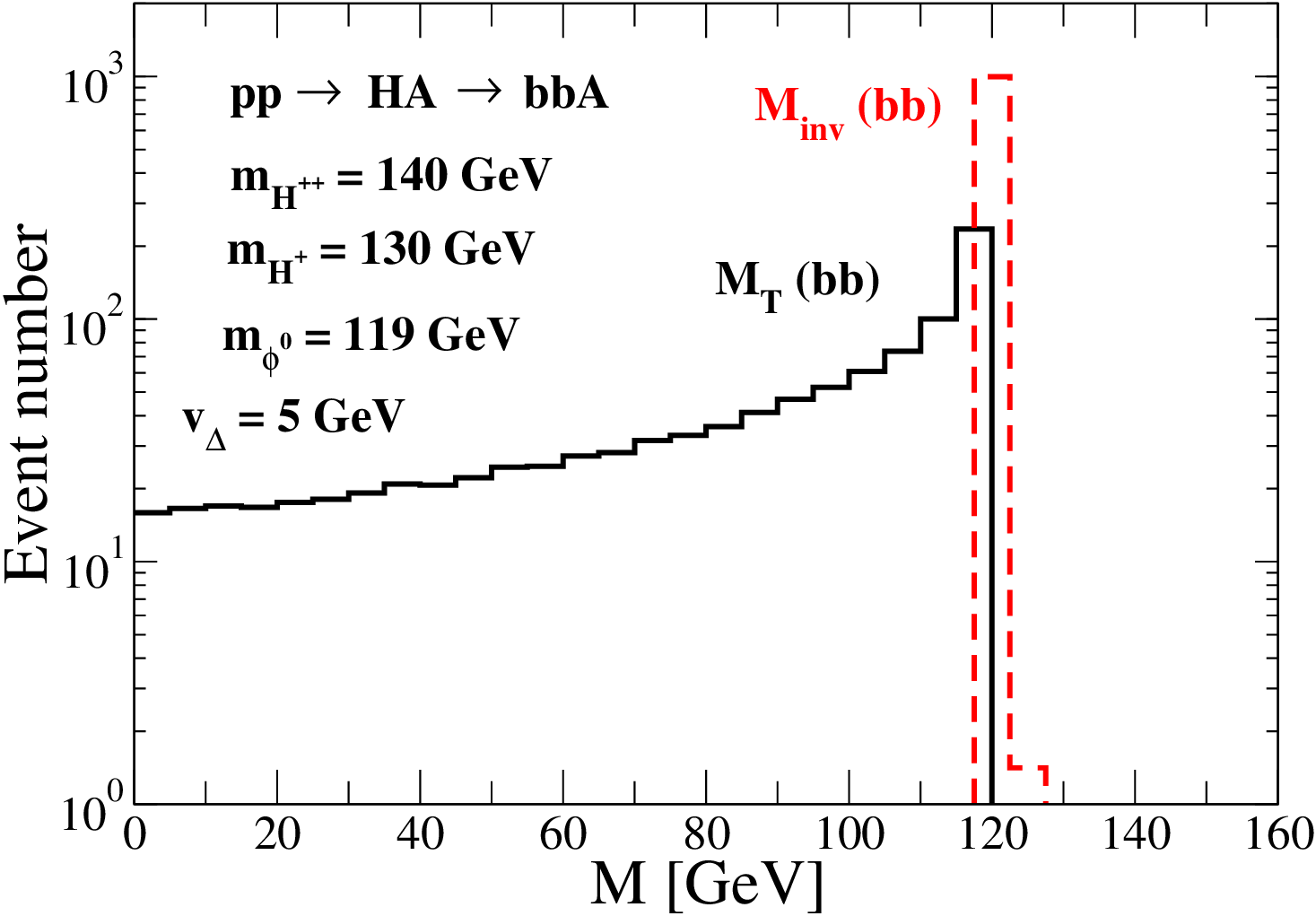}
\caption{The transvers mass distributions for each system in Scenario (1a). 
The total event number is assumed to be 1000. 
In the bottom-right figure, 
the horizontal axis $M$ represents the 
transverse mass distribution for the $b\bar{b}$ system $M_T(bb)$ (solid) or the
invariant mass distribution for the $b\bar{b}$ system $M_{\text{inv}}(bb)$ (dashed). }
\label{mt2}
\end{center}
\end{figure}

\begin{figure}[t]
\begin{center}
\includegraphics[width=50mm]{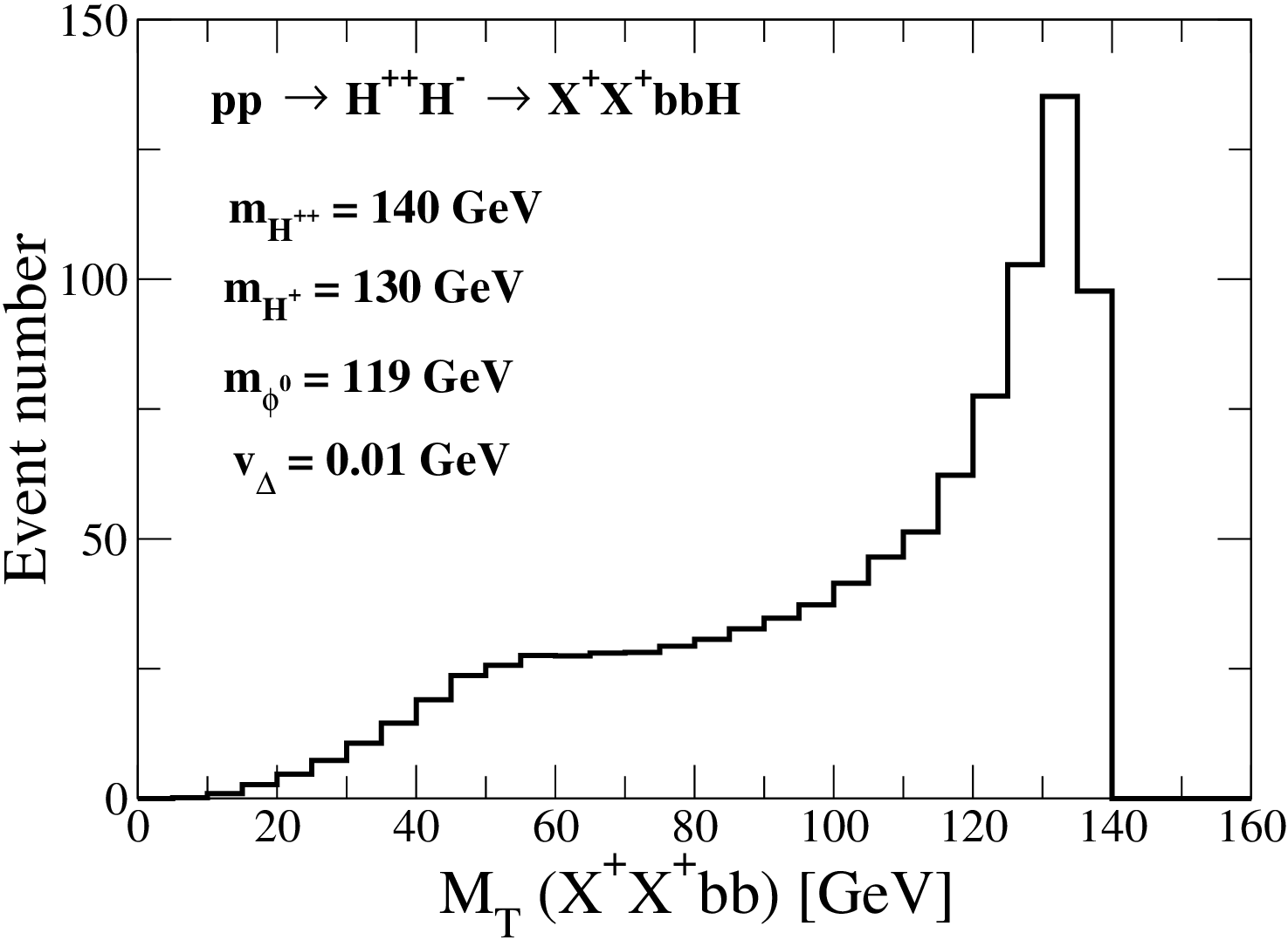}\hspace{3mm}
\includegraphics[width=50mm]{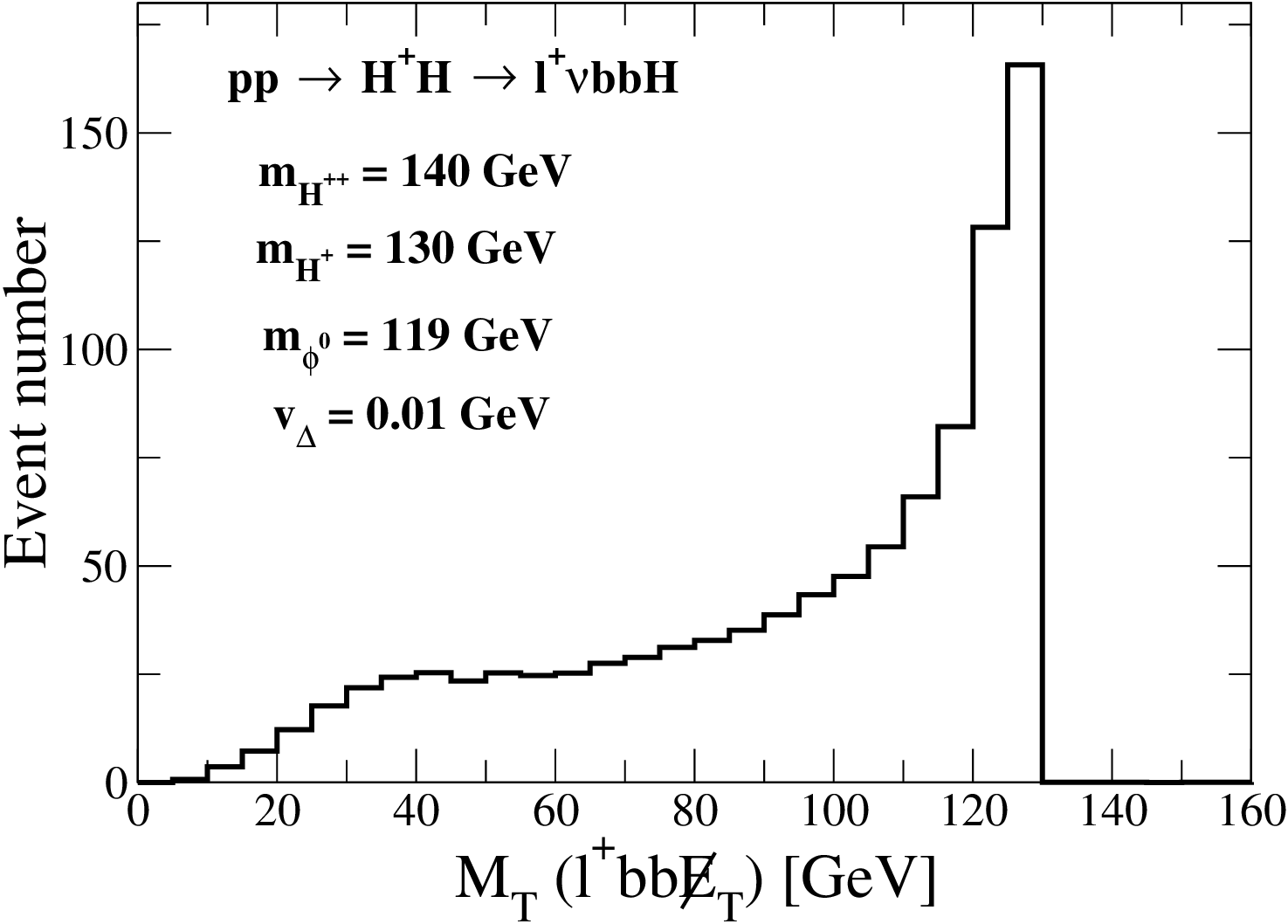}\hspace{3mm}
\includegraphics[width=50mm]{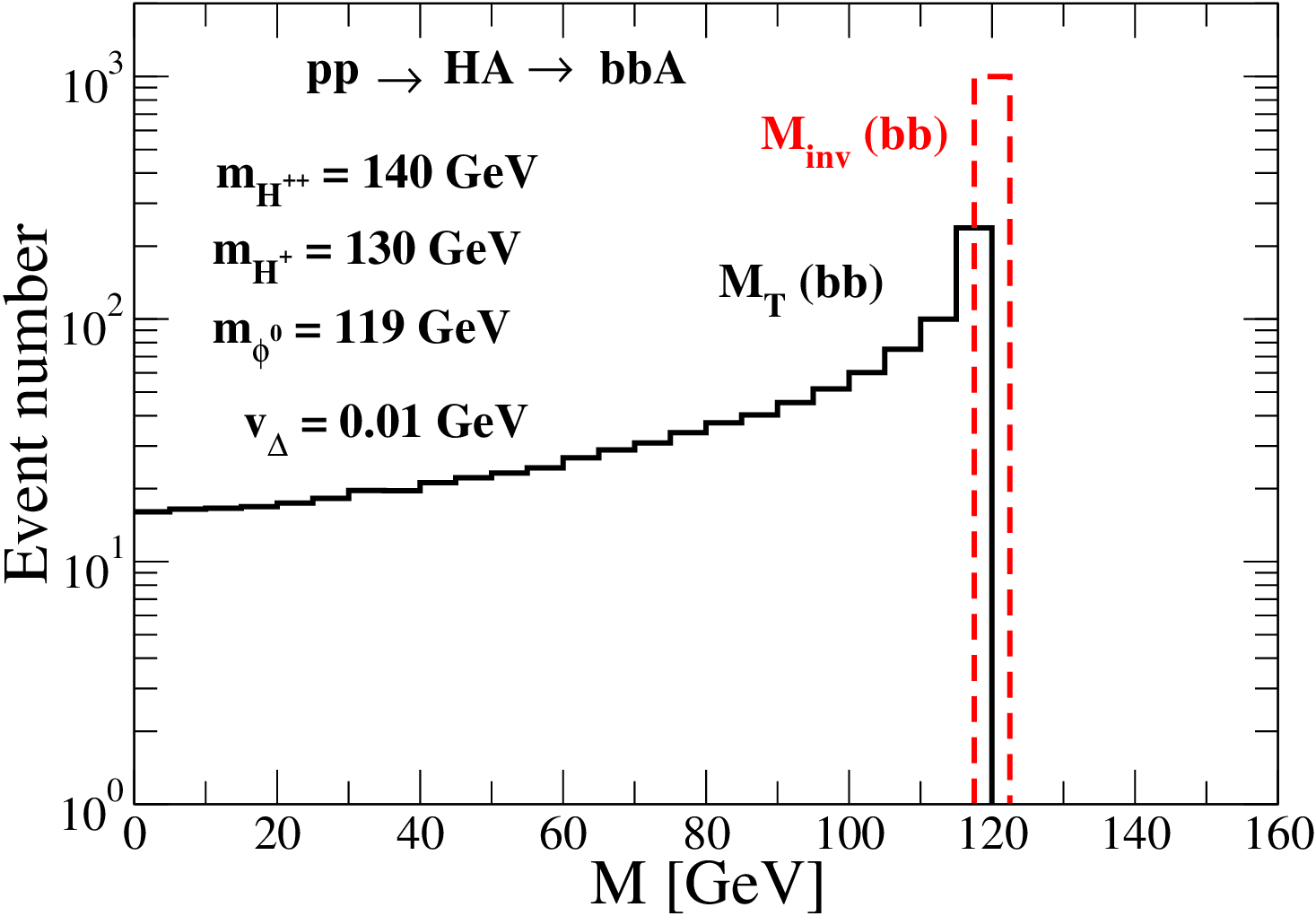}
\caption{The transvers mass distributions for each system in Scenario (1b). 
The total event number is assumed to be 1000. 
In the right figure, 
the horizontal axis $M$ represents the 
transverse mass distribution for the $b\bar{b}$ system $M_T(bb)$ (solid) or the 
invariant mass distribution for the $b\bar{b}$ system $M_{\text{inv}}(bb)$ (dashed). }
\label{mt1}
\end{center}
\end{figure}

\begin{description}
\item[Scenario (1a)];\\    
We can measure $m_{H^{++}}$ by observing 
the endpoint in the transverse mass distribution of the $\ell^+\ell^+ E_T\hspace{-4mm}/\hspace{3mm}$ system 
in the process $pp\to H^{++}H^-\to (W^{+*}W^+)(W^{-*}Z)\to (\ell^+\ell^+ E_T\hspace{-4mm}/\hspace{3mm})(jjjj)$, (FIG.~\ref{mt2} upper left). 
At the same time, we can also determine $m_{H^+}$ by measuring the 
endpoint in the transverse mass distribution of the 
$\ell^+ jjE_T\hspace{-4mm}/\hspace{3mm}$ system or the $\ell^+ E_T\hspace{-4mm}/\hspace{3mm}$ system 
in the process $pp\to H^{+}\phi^0\to (W^{+*}Z)(b\bar{b})\to (\ell^+ jjE_T\hspace{-4mm}/\hspace{3mm})(j_bj_b)$ or 
$pp\to H^{+}\phi^0\to (\tau^+\nu)(b\bar{b})\to (\ell^+E_T\hspace{-4mm}/\hspace{3mm})(j_bj_b)$, 
(FIG.~\ref{mt2} upper right and lower left). 
In addition, $m_{\phi^0}$ can be determined by using 
the invariant mass distribution 
or by observing the endpoint in the transverse mass distribution 
of the $b\bar{b}$ system in 
the process $pp\to HA\to (b\bar{b})(b\bar{b})\to (j_bj_b)(j_bj_b)$, (FIG.~\ref{mt2} lower right).
\item[Scenario (1b)];\\
We can determine $m_{H^{++}}$ by measuring 
the endpoint in the transverse mass distribution of the $\ell^+\ell^+j_bj_bE_T\hspace{-4mm}/\hspace{3mm}$ system 
in the process $pp\to H^{++}H^-\to (W^{+*}W^{+*}b\bar{b})(W^{-*}b\bar{b})\to (\ell^+\ell^+j_bj_bE_T\hspace{-4mm}/\hspace{3mm})(jjj_bj_b)$, 
(FIG.~\ref{mt1} left). 
Analysing the transverse mass distribution for the $\ell^+\ell^+j_bj_bE_T\hspace{-4mm}/\hspace{3mm}$ system,  
we treat that a lepton pair $\ell^+\nu$ from $W^{+*}$ as one massless fermion as represented $X^+$ in FIG.~\ref{mt1}. 
This procedure is justified since the angle between $\ell^+$ and $\nu$ is distributed almost around $0^\circ$. 
We can also determine $m_{H^+}$ 
by measuring the endpoint in the transverse mass distribution of the $\ell^+ j_bj_bE_T\hspace{-4mm}/\hspace{3mm}$ system 
in the process $pp\to H^{+}\phi^0\to (W^{+*}b\bar{b})(b\bar{b})\to (\ell^+ j_bj_bE_T\hspace{-4mm}/\hspace{3mm})(j_bj_b)$, (FIG.~\ref{mt1} center). 
In addition, 
$m_{\phi^0}$ can be reconstructed 
by measuring the invariant mass distribution of the $b\bar{b}$ system 
and by observing the endpoint of the transverse mass distribution of the $b\bar{b}$ system 
in the process $pp\to HA\to (b\bar{b})(b\bar{b})\to (j_bj_b)(j_bj_b)$ (FIG.~\ref{mt1} right). 
\item[Scenario (1c)];\\
The final state of the decay of the triplet-like scalar bosons always include neutrinos, 
so that the reconstruction of 
the masses of the triplet-like scalar bosons would be challenging.
\item[Scenario (1d)];\\
This scenario is already excluded from the direct search results at the LHC for the processes of 
$pp\to H^{++}H^{--}(H^{\pm\pm}H^\mp)\to \ell^+\ell^+\ell^-\ell^-(\ell^\pm\ell^\pm\ell^\mp\nu)$. 
\end{description}

In TABLE~\ref{t3}, processes which can use the reconstruction of the masses of the triplet-like scalar bosons 
are summarized in each scenario. 
The cross sections for the final states of each process are also listed. 
In the case of Set~2, the masses of the triplet like scalar bosons may be able to reconstruct in the similar way to the case 
of Set~1. 
Thus, we show only the signal cross sections for the final states for Set~2 in TABLE~\ref{t4}.
\begin{table}[t]
\begin{center}
{\renewcommand\arraystretch{1.3}
\begin{tabular}{|c||c|c|c|}\hline
& $m_{H^{++}}$ & $m_{H^+}$ & $m_H/m_A$\\\hline\hline
(1a) &
{\small$pp\to H^{++}H^-\to (\ell^+\ell^+E_T\hspace{-4mm}/\hspace{3mm})(jjjj)$ }
&{\small$pp\to H^{+}H\to (\ell^+jjE_T\hspace{-4mm}/\hspace{3mm})(j_bj_b)$ } &
{\small$pp\to HA\to (j_bj_b)(j_bj_b)$ }\\
&{\small [2.8 fb] (0.95 fb)} & {\small[11 fb] (3.8 fb)} & {\small [270 fb] (95 fb)}
\\
&
&{\small$pp\to H^{+}H\to (\ell^+E_T\hspace{-4mm}/\hspace{3mm})(j_bj_b)$ }  &
{\small$pp\to H^{+}H\to (\ell^+E_T\hspace{-4mm}/\hspace{3mm})(j_bj_b)$ }\\
&
& {\small[9.3 fb] (3.3 fb)} & {\small[9.3 fb] (3.3 fb)}
\\\hline
(1b) &{\small$pp\to H^{++}H^-\to (\ell^+\ell^+j_bj_bE_T\hspace{-4mm}/\hspace{3mm})(jjj_bj_b)$}&{\small $pp\to H^+H\to (\ell^+j_bj_bE_T\hspace{-4mm}/\hspace{3mm})(j_bj_b)$} &{\small$pp\to HA\to (j_bj_b)(j_bj_b)$} \\
&{\small [8.4 fb] (2.9 fb)} & {\small[36 fb] (13 fb)} & {\small[270 fb] (95 fb)}\\
&
& &{\small$pp\to H^+H\to (\ell^+j_bj_bE_T\hspace{-4mm}/\hspace{3mm})(j_bj_b)$} \\
&
&  & {\small[36 fb] (13 fb)}\\\hline
(1c) &\multicolumn{3}{c|}{Challenging}\\\hline
(1d) &\multicolumn{3}{c|}{Excluded}\\\hline
\end{tabular}}
\caption{
The processes which can be used to reconstruct the masses of the triplet-like scalar bosons 
are summarized. 
The numbers in [ ] and ( ) represent the cross section for the final state of the process at $\sqrt{s}=14$ TeV and $\sqrt{s}=7$ TeV, 
respectively, for Set~1.   
The values of the decay branching ratios of the triplet-like scalar bosons are listed in TABLE~\ref{tdecay1}. 
In this table, the $b$-tagging efficiency is assumed to be 100 \%. }
\label{t3}
\end{center}
\end{table}

\begin{table}[t]
\begin{center}
{\renewcommand\arraystretch{1.3}
\begin{tabular}{|c||c|c|c|}\hline
& $m_{H^{++}}$ & $m_{H^+}$ & $m_H/m_A$\\\hline\hline
(2a) &
{\small$pp\to H^{++}H^-\to (\ell^+\ell^+E_T\hspace{-4mm}/\hspace{3mm})(jjj_bj_b)$ }
&{\small$pp\to H^{+}H\to (\ell^+j_bj_bE_T\hspace{-4mm}/\hspace{3mm})(j_bj_b)$ } &
{\small$pp\to HA\to (j_bj_b)(j_bj_b)$ }\\
&{\small [2.7 fb] (0.84 fb)} & {\small[21 fb] (6.9 fb)} & {\small [230 fb] (76 fb)}
\\\hline
(2b) &{\small$pp\to H^{++}H^-\to (\ell^+\ell^+j_bj_bE_T\hspace{-4mm}/\hspace{3mm})(jjj_bj_b)$}&{\small $pp\to H^+H\to (\ell^+j_bj_bE_T\hspace{-4mm}/\hspace{3mm})(j_bj_b)$} &{\small$pp\to HA\to (j_bj_b)(j_bj_b)$} \\
&{\small [3.2 fb] (0.99 fb)} & {\small[22 fb] (7.2 fb)} & {\small[230 fb] (76 fb)}\\\hline
(2c) &\multicolumn{3}{c|}{Challenging}\\\hline
(2d) &\multicolumn{3}{c|}{Excluded}\\\hline
\end{tabular}}
\caption{
The processes which can be used to reconstruct the masses of the triplet-like scalar bosons 
are summarized. 
The numbers in [ ] and ( ) represent the cross section for the final state of the process at $\sqrt{s}=14$ TeV and $\sqrt{s}=7$ TeV, 
respectively, for Set~2.   
The values of the decay branching ratios of the triplet-like scalar bosons are listed in TABLE~\ref{tdecay2}. 
In this table, the $b$-tagging efficiency is assumed to be 100 \%.  }
\label{t4}
\end{center}
\end{table}

\section{Discussions}
 
We give comments on the discrimination of the model from the others which 
contain doubly-charged scalar bosons such as that from $Y=2$ singlet scalar fields and/or $Y=3/2$ doublet scalar fields. 
First, doubly-charged scalar bosons from singlet fields 
appear in the Zee-Babu model~\cite{Zee-Babu} which generates neutrino masses at the 2-loop level.  
The doubly-charged scalar bosons from this model 
do not couple to $W$ boson. 
Thus, it may be distinguished by the production process for 
doubly-charged scalar bosons associated with singly-charged scalar bosons. 
Second, we consider the discrimination of the HTM from the model with the $Y=3/2$ doublet $H_{3/2}=(H_{3/2}^{++},H_{3/2}^+)$. 
The doubly-charged component field $H^{++}_{3/2}$ decays into 
$H^+_{3/2}W^{+(*)}$ because $H_{3/2}$ does not receive the VEV as discussed in Ref.~\cite{AKY}
\footnote{If higher order operators are introduced as discussed in Refs.~\cite{gunion,su}, 
$H^{++}_{3/2}$ can decay into the same sign dilepton. }.  
$H^+_{3/2}$ decays into $\tau^+\nu$ or $cs$ via the mixing with 
the singly-charged scalar boson from the $Y=1/2$ doublet fields
\footnote{$H^+_{3/2}$ can decay into the SM particles if the model has two Higgs doublet fields with $Y=1/2$ at least.} 
while $H^+$ decays into $\phi^0W^{+(*)}$ in the HTM. 
Therefore, we can distinguish these models because the final state is different.  
Finally, the singly-charged Higgs boson in the HTM can also be discriminated from 
that in the two Higgs doublet model (THDM).  
In both the models $H^+$ is produced via $pp\to W^{+*}\to H^+A$ $(H^+H)$. 
In the THDM with the type II Yukawa interaction including the minimal supersymmetric standard model
\footnote{The mass of $H^\pm$ with $\mathcal{O}$(100) GeV is highly constrained by the $b\to s\gamma$ experiments~\cite{bsg} in the 
general type II THDM. }, 
$H^+$ decays into $\tau^+\nu$ or $t\bar{b}$ depending on the mass 
of the $H^+$ while 
$H$ or $A$ decays into $b\bar{b}$ as long as the masses are not too heavy.  
Therefore, the final state of the $pp\to H^+A$ is $\tau^+\nu b\bar{b}$ or $W^+b\bar{b}b\bar{b}$ in the type II THDM~\cite{Cao:2003tr}. 
Although the latter final state is the same as that in the HTM with $\xi> 0$ we may be 
able to distinguish these models by reconstructing the top quark in the $W^+ b$ system. 
In the THDM with the type X Yukawa interaction (the lepton specific THDM), 
extra neutral Higgs bosons decay into $\tau^+\tau^-$ instead of $b\bar{b}$~\cite{typeX}, which 
is different from the HTM, and we would be able to separate the models.

We have discussed the case of light triplet-like scalar bosons with these masses of $\mathcal{O}$ (100) GeV.  
Here, we comment on a rather heavy triplet-like scalar bosons case, 
e.g., $m_{H^{++}}$= 320 GeV, $m_{H^+}$= 310 GeV and $m_{\phi^0}$= 300 GeV. 
The biggest change should be in the decay of the neutral scalar bosons; i.e.,  
$H$ ($A$) decays into $hh$ ($hZ$) when $v_\Delta \gtrsim 10^{-3}$ GeV (see FIG.~\ref{fig3}). 
Decay of $H^{++}$ is almost the same as in the case of the light triplet-like scalar bosons case (see FIG.~\ref{fig1}). 
$H^+$ can decay into $t\bar{b}$ instead of $\tau^+\nu$ (see FIG.~\ref{fig2}). 
Since decay modes of the neutral scalar bosons change, 
final states via $H$ or $A$ include more jets. 
At the same time, the production cross sections for the Drell-Yan processes decrease in this case. 
For instance, 
the cross sections for $pp\to H^{++}H^-$, $pp\to H^+\phi^0$ and $pp\to HA$ 
are 
15 fb (3.4 fb), 7.7 fb (1.7 fb) and 5.6 fb (1.1 fb), respectively at $\sqrt{s}=$ 14 TeV (7 TeV).  Therefore, 
measurement the masses of the triplet-like scalar bosons is rather challenging, especially 
when $H^{++}$ decays into $H^+W^{+(*)}$. 

In this paper, we only have discussed the signal processes and 
we have not discussed backgrounds against the signal. 
The background analysis is beyond the scope in this paper. 
It would be expected that the backgrounds can be reduced after the appropriate kinematic cuts. 
For example, in the case of Scenario (1b), 
the main background against the signal  
$pp\to H^{++}H^-\to (\ell^+\ell^+b\bar{b}E_T\hspace{-4mm}/\hspace{2.5mm})(jjb\bar{b})$ 
may come from $t\bar{t}W^+W^-$. 
Typically, the cross section for this background is $\mathcal{O}$(1) pb at $\sqrt{s}=14$ TeV. 
In the case where one of the $W$ boson from the top quark decays hadronicaly; i.e., 
$t\bar{t}W^+W^-\to (b\ell^+\nu)(\bar{b}jj)(\ell^+\nu)(jj)$, 
the cross section for the final state of the background would be $\mathcal{O}$(10) fb. 
On the other hand, the cross section for the final state of the signal 
$pp\to H^{++}H^-\to (\ell^+\ell^+b\bar{b}E_T\hspace{-4mm}/\hspace{2.5mm})(jjb\bar{b})$ 
is 8.4 fb at $\sqrt{s}=14$ TeV. 
Although the cross sections for the signal and the background 
are comparable at this stage, 
the background can be further reduced by 
reconstructing the top quarks by using the invariant mass distribution 
of the $bjj$ system and the endpoint in the transverse mass distribution 
of the $b\ell E_T\hspace{-4mm}/\hspace{2.5mm}$ system. 
In the case where both the $W$ bosons from the top quarks decay leptonicaly; i.e.,   
$t\bar{t}W^+W^-\to (b\ell^+\nu)(\bar{b}\ell^-\nu)(jj)(jj)$, 
the reconstruction of the top quarks would be difficult. 
However, by using the electric charge identification for leptons,  
the background and the signal can be further separated. 
The main backgrounds against 
the signal $pp\to H^+\phi^0\to (\ell^+b\bar{b}E_T\hspace{-4mm}/\hspace{2.5mm})(b\bar{b})$ 
may come from $t\bar{t}$ and $t\bar{t}\gamma^*$/$t\bar{t}g^*$ whose cross sections would be  
$\mathcal{O}$(10) pb and $\mathcal{O}$(1) pb at $\sqrt{s}=14$ TeV, respectively. 
The cross section for the final state of the signal 
$pp\to H^+\phi^0\to (\ell^+b\bar{b}E_T\hspace{-4mm}/\hspace{2.5mm})(b\bar{b})$ 
is 36 fb at $\sqrt{s}=14$ TeV. 
For the background of $t\bar{t}\to (b\ell^+\nu)(\bar{b}jj)$, 
the top quarks can be reconstructed similarly in the case of the 
background of 
$t\bar{t}W^+W^-\to (b\ell^+\nu)(\bar{b}jj)(\ell^+\nu)(jj)$, 
so that this background would be separated. 
Next, the process $t\bar{t}\gamma^*$/$t\bar{t}g^*\to 
(b\ell^+\nu)(\bar{b}\ell^-\nu)(jj)$ can be a background 
if one of the charged lepton is miss identified. 
The cross section for the final state of the background 
would be $\mathcal{O}$(10) fb when the miss-identification rate for a charged lepton 
is assumed to be $10\%$. 
The cross sections for the signal and the background 
are comparable at this stage. 
In this case, although we may not be able to use the top quark reconstruction, 
by using the $b$-tagging and the cuts for the low energy jet, 
the background would be expected to be reduced. 
However, it goes without saying that the 
detector level simulation is necessary to clarify the feasibility of the signal. 
This would be a future task. 

\section{Conclusions}

In the HTM, a characteristic mass spectrum 
$\xi = m_{H^{++}}^2-m_{H^+}^2\simeq m_{H^+}^2-m_{\phi^0}^2$ is predicted when $v_\Delta \ll v$. 
Therefore, by measuring this mass spectrum of the triplet-like scalar bosons, 
the model can be tested at the LHC.  
We have investigated the collider signature in the HTM with $\xi > 0$ at the LHC.  
In this case, $H^{++}$ is the heaviest of all the triplet-like scalar bosons. 
When $v_\Delta > 10^{-4}-10^{-3}$ GeV, $H^{++}$ does not decay into the same sign 
dilepton so that the limit of the mass of $H^{++}$ from the recent results at the LHC 
cannot be applied. 
We thus mainly have discussed the case of light triplet-like scalar bosons 
whose masses are of $\mathcal{O}$(100) GeV. 
In such a case, triplet-like scalar bosons mainly decay into 
$H^{++}\to H^+W^{+(*)}$, $H^+\to \phi^0 W^{+(*)}$ and $\phi^0\to b\bar{b}$. 
We have found that all the masses of the triplet-like scalar bosons  
may be able to be reconstructed by measuring the endpoint in 
the transverse mass distribution and the invariant mass distribution of 
the systems which are produced via the decay of the triplet-like scalar bosons. 

Detector level simulation should be necessary to clarify the feasibility of 
measuring the masses of the triplet-like scalar bosons. 
\\\\
$Acknowledgments$
The authors would like to thank Hiroaki Sugiyama for a useful comment. 
The work of M.A. was supported in part by Grant-in-Aid for Scientific Research, 
No. 22740137. 
The work of S.K. was supported in part by Grant-in-Aid for Scientific Research, Nos. 22244031 and 23104006. 
K.Y. was supported by Japan Society for the Promotion of Science.

\appendix
\section{Decay rates of the triplet-like scalar bosons}
In this Appendix, we list the formulae of decay rates for $H^{\pm\pm}$, $H^\pm$, $H$ and $A$ in order. 
\subsection{Decay rates of $H^{\pm\pm}$}
The decay rates for $H^{\pm\pm}$ can be evaluated as 
\begin{align}
\Gamma(H^{\pm\pm} \to \ell_i^\pm \ell_j^\pm)
&=S_{ij}|h_{ij}|^2\frac{m_{H^{++}}}{4\pi}\left(1-\frac{m_i^2}{m_{H^{++}}^2}-\frac{m_j^2}{m_{H^{++}}^2}\right)\left[\lambda\left(\frac{m_i^2}{m_{H^{++}}^2},\frac{m_j^2}{m_{H^{++}}^2}\right)\right]^{1/2},\label{a1}\\
\Gamma(H^{\pm\pm} \to W^\pm W^\pm)&=\frac{g^4v_\Delta^2m_{H^{++}}^3}{16\pi m_W^4}\left(\frac{3m_W^4}{m_{H^{++}}^4}-\frac{m_W^2}{m_{H^{++}}^2}+\frac{1}{4}\right)\beta\left(\frac{m_W^2}{m_{H^{++}}^2}\right),\\
\Gamma(H^{\pm\pm} \to H^\pm W^\pm )&=\frac{g^2m_{H^{++}}^3}{16\pi m_W^2}\cos^2\beta_\pm\left[\lambda\left(\frac{m_W^2}{m_{H^{++}}^2},\frac{m_{H^+}^2}{m_{H^{++}}^2}\right)\right]^{3/2},\\
\Gamma(H^{\pm\pm} \to W^\pm W^{\pm *}) 
&=\frac{3g^6m_{H^{++}}}{512\pi^3}\frac{v_{\Delta}^2}{m_W^2}F\left(\frac{m_W^2}{m_{H^{++}}^2}\right),\\
\Gamma(H^{\pm\pm} \to H^\pm W^{\pm *})&=
\frac{9g^4m_{H^{++}}}{128\pi^3}\cos^2\beta_\pm G\left(\frac{m_{H^+}^2}{m_{H^{++}}^2},\frac{m_W^2}{m_{H^{++}}^2}\right),
\end{align}
where $m_i$ is the lepton mass ($i=e,\mu$ or $\tau$) and $S_{ij}=1$, $(1/2)$ for $i\neq j$, $(i=j)$. 
The functions of $\lambda(x,y)$, $\beta(x)$, $F(x)$ and $G(x,y)$ are 
\begin{align}
\lambda(x,y)&=1+x^2+y^2-2xy-2x-2y,\\
\beta(x)&=\sqrt{\lambda(x,x)}=\sqrt{1-4x},\\
F(x)&=-|1-x|\left(\frac{47}{2}x-\frac{13}{2}+\frac{1}{x}\right)+3(1-6x+4x^2)|\log \sqrt{x}|+\frac{3(1-8x+20x^2)}{\sqrt{4x-1}}\arccos\left(\frac{3x-1}{2x^{3/2}}\right), \\
G(x,y)&=\frac{1}{12y}\Bigg\{2\left(-1+x\right)^3-9\left(-1+x^2\right)y+6\left(-1+x\right)y^2\notag\\
&+6\left(1+x-y\right)y\sqrt{-\lambda(x,y)}\left[\arctan\left(\frac{-1+x-y}{\sqrt{-\lambda(x,y)}}\right)+\arctan\left(\frac{-1+x+y}{\sqrt{-\lambda(x,y)}}\right)\right]\notag\\
&-3\left[1+\left(x-y\right)^2-2y\right]y\log x\Bigg\}.  \label{g_func}
\end{align}
Although the expression in Eq.~(\ref{g_func}) is different from that in Ref.~\cite{Djouadi:1995gv}, 
we have confirmed that the numerical value by using Eq.~(\ref{g_func}) coincides with 
that by using CalcHEP.

\subsection{Decay rates of $H^\pm$}
The decay rates for $H^{\pm}$ can be evaluated as 
\begin{align}
\Gamma(H^\pm \to q\bar{q}')&=\frac{3m_{H^+}^3}{8\pi v^2}\sin^2\beta_\pm 
\left[\left(\frac{m_q^2}{m_{H^+}^2}+\frac{m_{q'}^2}{m_{H^+}^2}\right)\left(1-\frac{m_q^2}{m_{H^+}^2}-\frac{m_{q'}^2}{m_{H^+}^2}\right)-4\frac{m_q^2}{m_{H^+}^2}\frac{m_{q'}^2}{m_{H^+}^2}\right]\notag\\
&\hspace{23mm}\times\left[\lambda\left(\frac{m_q^2}{m_{H^+}^2},\frac{m_{q'}^2}{m_{H^+}^2}\right)\right]^{1/2},\\
\Gamma(H^\pm \to \ell_i^\pm\nu_j)&=\delta_{ij}\frac{m_i^2m_{H^+}}{8\pi v^2}\sin^2\beta_\pm \left(1-\frac{m_i^2}{m_{H^+}^2}\right)^2+|h_{ij}|^2\frac{m_{H^+}}{8\pi}\cos^2\beta_\pm\left(1-\frac{m_i^2}{m_{H^+}^2}\right)^2,\\
\Gamma(H^\pm \to W^\pm Z)
& =\frac{g^2g_Z^2}{32\pi m_{H^+}}v_\Delta^2\cos^2\beta_\pm\left[\lambda\left(\frac{m_W^2}{m_{H^+}^2},\frac{m_Z^2}{m_{H^+}^2}\right)\right]^{1/2}\left[2+\frac{m_{H^+}^4}{4m_W^2m_Z^2}\left(1-\frac{m_W^2}{m_{H^+}^2}-\frac{m_Z^2}{m_{H^+}^2}\right)^2\right],\\
\Gamma(H^\pm \to W^\pm Z^* )&=\frac{3g^2g_Z^4}{1024\pi^3m_{H^+}}v_\Delta^2\cos^2\beta_\pm H\left(\frac{m_W^2}{m_{H^+}^2},\frac{m_Z^2}{m_{H^+}^2}\right)\left(7-\frac{40}{3}\sin^2\theta_W+\frac{160}{9}\sin^4\theta_W\right),\\
\Gamma(H^\pm \to W^{\pm *} Z)&=\frac{9g^4g_Z^2}{512\pi^3 m_{H^+}}v_\Delta^2\cos^2\beta_\pm H\left(\frac{m_Z^2}{m_{H^+}^2},\frac{m_W^2}{m_{H^+}^2}\right),\\
\Gamma(H^\pm \to \hat{\varphi} W^\pm ) & =\frac{g^2m_{H^+}^3}{64\pi^2m_W^2}\xi_{H^+ W^- \hat{\varphi}}^2\left[\lambda\left(\frac{m_W^2}{m_{H^+}^2},\frac{m_{\hat{\varphi}}^2}{m_{H^+}^2}\right)\right]^{3/2},\label{a2}\\
\Gamma(H^\pm \to \hat{\varphi} W^{\pm *}  ) & =\frac{9g^4m_{H^+}}{512\pi^3}\xi_{H^+ W^- \hat{\varphi}}^2G\left(\frac{m_{\hat{\varphi}}^2}{m_{H^+}^2},\frac{m_W^2}{m_{H^+}^2}\right),\label{a3}
\end{align}
where $g_Z=g/\cos\theta_W$ with $\theta_W$ is the weak angle. 
The function $H(x,y)$ is
\begin{align}
H(x,y)&=\frac{\arctan\left[\frac{1-x+y}{\sqrt{-\lambda(x,y)}}\right]
+\arctan\left[\frac{1-x-y}{\sqrt{-\lambda(x,y)}}\right]}{4x \sqrt{-\lambda(x,y)}}
\Big[-3x^3+(9y+7)x^2-5(1-y)^2x+(1-y)^3\Big] \notag\\
&+\frac{1}{24xy}\Bigg\{(-1+x)[6y^2+y(39x-9)+2(1-x)^2]-3y[y^2+2y(3x-1)-x(3x+4)+1]\log x\Bigg\}. \label{hfunc}
\end{align}
We have confirmed that the numerical value by using Eq.~(\ref{hfunc}) coincides with 
that by using CalcHEP. 
In Eq.~(\ref{a2}) and Eq.~(\ref{a3}), 
$\hat{\varphi}$ denotes $h$, $H$ or $A$ and $\xi_{H^+W^- \hat{\varphi}}$ is expressed as 
\begin{align}
\xi_{H^+W^- h}=\cos\alpha\sin\beta_\pm-\sqrt{2}\sin\alpha\cos\beta_\pm,\notag\\
\xi_{H^+W^- H}=\sin\alpha\sin\beta_\pm+\sqrt{2}\cos\alpha\cos\beta_\pm,\notag\\
\xi_{H^+W^- A}=\sin\beta_0\sin\beta_\pm+\sqrt{2}\cos\beta_0\cos\beta_\pm.
\end{align}

\subsection{Decay rates of $H$}
The decay rates for $H$ can be evaluated as 
\begin{align}
\Gamma(H\to f\bar{f})&=\frac{N_c^fm_f^2m_H}{8\pi v^2}\sin^2\alpha\left[\beta\left(\frac{m_f^2}{m_H^2}\right)\right]^3,\\
\Gamma(H \to \nu \nu)&=\Gamma(H \to \nu^c \bar{\nu})+\Gamma(H \to \bar{\nu}^c \nu)
=\sum_{i,j=1}^3S_{ij}|h_{ij}|^2\frac{m_H}{4\pi}\cos^2\alpha,\\
\Gamma(H\to W^+W^-)&=\frac{g^4 m_H^3}{16\pi m_W^4}\left(\frac{v}{2}\sin\alpha -v_\Delta \cos\alpha\right)^2\left(\frac{1}{4}-\frac{m_W^2}{m_H^2}+\frac{3m_W^4}{m_H^4}\right)\beta\left(\frac{m_W^2}{m_H^2}\right),\\
\Gamma(H\to ZZ)&=\frac{g_Z^4 m_H^3}{32\pi m_Z^4}\left(\frac{v}{2}\sin\alpha -2v_\Delta \cos\alpha\right)^2\left(\frac{1}{4}-\frac{m_Z^2}{m_H^2}+\frac{3m_Z^4}{m_H^4}\right)\beta\left(\frac{m_Z^2}{m_H^2}\right),\\
\Gamma(H\to WW^*)&=\frac{3 g^6 m_H}{512\pi^3}\frac{(\frac{v}{2}\sin\alpha-v_\Delta \cos\alpha)^2}{m_W^2}F\left(\frac{m_W^2}{m_H^2}\right),\\
\Gamma(H\to ZZ^*)&=\frac{g_Z^6 m_H}{2048\pi^3}\frac{(\frac{v}{2}\sin\alpha-2v_\Delta \cos\alpha)^2}{m_Z^2}\left(7-\frac{40}{3}\sin^2\theta_W+\frac{160}{9}\sin^4\theta_W\right)F\left(\frac{m_Z^2}{m_H^2}\right),\\
\Gamma(H\to hh)&=\frac{\lambda_{Hhh}^2}{8\pi m_H}\beta\left(\frac{m_h^2}{m_H^2}\right),
\end{align}
where 
\begin{align}
\lambda_{Hhh}&=\frac{1}{4v^2}\Big\{2v_\Delta\left[-2M_\Delta^2+v^2(\lambda_4+\lambda_5)\right]\cos^3\alpha+v^3\left[-12\lambda_1+4(\lambda_4+\lambda_5)\right]\cos^2\alpha\sin\alpha \notag\\
&+4v_\Delta\left[2M_\Delta^2+v^2(3\lambda_2+3\lambda_3-\lambda_4-\lambda_5)\right]\cos\alpha\sin^2\alpha-2v^3(\lambda_4+\lambda_5)\sin^3\alpha\Big\}\notag\\
&\simeq \frac{1}{4v^2}\Big\{2v_\Delta\left[-2M_\Delta^2+v^2(\lambda_4+\lambda_5)\right]\cos^3\alpha+v^3\left[-12\lambda_1+4(\lambda_4+\lambda_5)\right]\cos^2\alpha\sin\alpha\Big\}, 
\end{align}
and $N_c^f$ is the color factor with $N_c^q=3$, $N_c^\ell=1$. 
\subsection{Decay rates of $A$}
The decay rates for $H$ can be evaluated as 
\begin{align}
\Gamma(A\to f\bar{f})&=\sin^2\beta_0\frac{N_c^f m_f^2 m_A}{8\pi v^2}\beta\left(\frac{m_f^2}{m_A^2}\right),\\
\Gamma(A\to \nu\nu)&=\Gamma(A\to \nu^c\bar{\nu})+\Gamma(A\to \bar{\nu}^c\nu)=\sum_{i,j=1}^3S_{ij}|h_{ij}|^2\frac{m_A}{4\pi}\cos^2\beta_0,\\
\Gamma(A\to hZ)&=\frac{g_Z^2m_A^3}{64\pi m_Z^2}(\cos\alpha\sin\beta_0-2\sin\alpha\cos\beta_0)^2\left[\lambda\left(\frac{m_h^2}{m_A^2},\frac{m_Z^2}{m_A^2}\right)\right]^{3/2},\\
\Gamma(A\to hZ^*)&=\frac{3g_Z^4}{1024\pi^3}(\cos\alpha\sin\beta_0-2\sin\alpha\cos\beta_0)^2m_AG\left(\frac{m_h^2}{m_A^2}, \frac{m_Z^2}{m_A^2}\right)\left(7-\frac{40}{3}\sin^2\theta_W+\frac{160}{9}\sin^4\theta_W\right).
\end{align}


\end{document}